\documentclass[11pt,a4paper,english]{article}
\usepackage[T1]{fontenc}
\usepackage[latin9]{inputenc}
\usepackage{babel}
\usepackage{amsmath}
\usepackage{amssymb}
\usepackage{amsfonts}
\usepackage{esint}
\usepackage{bbold}

\newcommand{\spec}{\mathrm{spec}}

\begin{document}

\title{Repeated quantum non-demolition measurements: convergence and continuous-time limit}

\author{Michel Bauer$^{\clubsuit}$%
\thanks{michel.bauer@cea.fr%
}, Tristan Benoist$^{\spadesuit}$%
\thanks{tristan.benoist@ens.fr%
}~ and Denis Bernard$^{\spadesuit}$%
\thanks{Member of CNRS; denis.bernard@ens.fr%
}}

\maketitle
\centerline{$^{\clubsuit}$ Institut de Physique Th\'eorique%
\footnote{CEA/DSM/IPhT, Unit\'e de recherche associ\'ee au CNRS%
} de Saclay, CEA-Saclay, France.} \centerline{$^{\spadesuit}$ Laboratoire
de Physique Th\'eorique de l'ENS,} \centerline{CNRS $\&$ Ecole Normale
Sup\'erieure de Paris, France.} \medskip{}

\begin{abstract}
We analyze general enough models of repeated indirect measurements in which a quantum system interacts repeatedly with randomly chosen probes on which Von Neumann direct measurements are performed. We prove, under suitable hypotheses, that the system state probability distribution converges after a large number of repeated indirect measurements, in a way compatible with quantum wave function collapse. Similarly  a modified version of the system density matrix converges. We show that the convergence is exponential with a rate given by some relevant mean relative entropies. We also prove that, under appropriate rescaling of the system and probe interactions, the state probability distribution and the system density matrix are solutions of stochastic differential equations modeling continuous-time quantum measurements. We analyze the large time convergence of these continuous-time processes and prove convergence.
\end{abstract}

\section{Introduction}
Repeated indirect quantum measurements aim at getting (partial) information on quantum systems with minimal impact on it. A possibility consists in repeating non-demolition measurements (QND). At each step, one lets the quantum system under study interact with another quantum system, called the probe, and performs a Von Neumann measurement\cite{Von Neumann} on this probe. Information on the quantum system is gained through intrication between the probe and the quantum system.
If one is aiming at progressively measuring a quantum observable, one has to make sure that a system prepared in one eigenstate of this observable remains in it after a cycle of intrication and direct measurement on the probe, and that the set of stable states, called pointer states, forms an orthonormal basis of the system Hilbert space. The experiment of ref.\cite{haroche}, in which repeated QND measurements is used to fix and measure the number of photons in a cavity without destroying them, illustrates this strategy.

Repeated indirect measurements were studied in ref.\cite{bauer_bernard}. There, the discussion was limited to QND measurements consisting of identical probes, interactions and measurements on the probe, and assuming a non-degeneracy condition. In the present article, we extend these results to cases where different indirect measurements (probes, interactions and direct measurements on probes) are used. We also study the degenerate case.

Let $Q_n(\cdot)$ be the pointer state distribution after the ${\rm n}^{\rm th}$ indirect measurement, that is $Q_n(\alpha)$ is the probability to find the system in the state $\alpha$ after n steps ($\alpha$ labels the pointer states), $\sum_\alpha Q_n(\alpha)=1$. As explain in section~\ref{sec:apparatus Bayes' law}, each cycle of indirect measurement updates the distribution through Bayes' law. The analysis of the distribution is reformulated as a problem in classical probability theory (with no quantum interference).

We shall prove that this sequence of distributions converges at large $n$, that is after a large -- strictly speaking, infinite -- number of QND measurements. If a non-degeneracy assumption is verified, the limit distribution is $Q(\alpha)=\delta_{\Upsilon,\alpha}$ for some random limit pointer state $\Upsilon$. This reflects the collapse of the system wave function. The convergence is exponential,
\[ Q_n(\alpha)\simeq {\rm const.}\ \exp(-n \overline{S}(\Upsilon|\alpha)),\quad {\rm for\ large}\ n, \alpha\not=\Upsilon,\]
with rate given by an appropriate relative entropy $\overline{S}(\Upsilon|\alpha)$ defined in eq.(\ref{meanS}).  
In probabilistic terms, the limit $Q(\alpha)$ possesses a natural interpretation as a Radon-Nikodym derivative and, $Q_n(\alpha)=\mathbb{E}(Q(\alpha)\vert {\cal F}_n)$ is a closed martingale with respect to an appropriate filtration, see section~\ref{sec:degeneracy}. As a consequence, we show that the expectations conditioned on the limit pointer state are identical to expectations starting from this same pointer state. That is: $\mathbb E_{\Upsilon}(\,\cdot\,)=\mathbb E(\,\cdot\,|\mathcal A)$ where $\mathcal A$ is the tail $\sigma$-algebra, the smallest one making the limit distribution measurable. See below for a precise definition of $\mathbb E_{\Upsilon}(\,\cdot\,)$. 

Convergence of $Q_n(\alpha)$ is also studied when the non-degeneracy hypothesis is not fulfilled. In this degenerate case, the limit pointer state distribution vanishes outside a random finite set of pointer states. The quantum system density matrix, when properly modified, also converges in the limit of infinite number of QND measurements. The limit density matrix then coincides with that predicted for degenerate Von Neumann measurements\cite{Von Neumann}, see section~\ref{sec:degeneracy}.

Of course repeated indirect measurements have already been studied in the physics literature, mostly through time continuous measurement formalisms -- as far as we know, little was done on the discrete setting as we do in the present paper. E. B. Davies\cite{davies} probably made the first rigorous approach to time continuous quantum measurement. This was later studied by N. Gisin\cite{gisin_84} and L. Diosi\cite{diosi_88} using the non linear Schr\"odinger equation. Simultaneously, A. Barchielli and V. P. Belavkin derived the equations governing continuous measurements in terms of instruments \cite{barchielli_belavkin_91}. They derived jump equations which, when properly rescaled, are equivalent to diffusive equations for continuous measurements. Another approach uses quantum stochastic differential equations and quantum filtering theory to obtain the so-called Belavkin equations \cite{barchielli_86,belavkin_89,belavkin_90,wiseman_94,bouten_04} and \cite{bouten_07}. More recently, C. Pellegrini derived Belavkin equations for continuous time measurements \cite{pellegrini_diffsaut,pellegrini_diff,pelegrini_saut} using discrete repeated indirect measurement models. The problem of convergence of quantum density matrix has also been analyzed within the time continuous measurement framework. In refs.\cite{belavkin_spin_92,belavkin_unsharp}, V. P. Belavkin showed the convergence of mixed states toward pure states. A derivation of wave function collapse from the non-linear stochastic Schr\"odinger equation has been presented in refs.\cite{adler_01,mabuchi_04}. It makes use of martingale theory as we do in the present paper.

In the following we also connect our discrete model to the time continuous measurement formalism. Taking the time continuous limit requires rescaling appropriately the interaction between the quantum system under study and the probes. In that sense, the time continuous model we consider is close to that of ref.\cite{pellegrini_diff} but our proofs are different and slightly more general. Our derivation is based on the convergence of some discrete counting processes -- related to the number of occurrences of outputs in the successive indirect measurements -- toward a time continuous Gaussian process. Under appropriate hypotheses, spelled out in section~\ref{sec:continuous time}, the pointer state distribution satisfies a random diffusive stochastic equation driven by Gaussian processes. Suppose that at each step the probe system is randomly selected (independently of the past history and with time independent probability, for simplicity) among a finite set ${\cal O}$ whose elements are indexed by $o\in {\cal O}$ and that the output measurements on the probe can take finite number of values indexed by $i\in {\rm spec}(o)$. Then, the pointer state probabilities $Q_t(\alpha)$ are time continuous martingales (with respect to an appropriate filtration) whose evolutions are governed by the non linear stochastic equations:
\[
 dQ_t(\alpha)=Q_t(\alpha)\sum_{(o,i)} \big(\Gamma^{(o)}(i|\alpha) - \langle \Gamma^{(o)}(i) \rangle_t\big)dX_t(o,i)
\]
where $X_t(o,i)$ are some centered Gaussian processes, $\Gamma^{(o)}(i|\alpha)$ are coding for the probability of output probe measurement $i$ within the probe system $o$ conditional on the quantum system be prepared in the state $\alpha$ and
$\langle \Gamma^{(o)}(i) \rangle_t=\sum_\alpha \Gamma^{(o)}(i|\alpha)
Q_t(\alpha)$. The pointer state distribution again converges as a
finite-dimensionnal bounded vector martingale. Under non-degeneracy assumptions, the limit distribution is again $Q(\alpha)=\delta_{\Upsilon,\alpha}$ and the convergence is still exponential with a rate given by the scaling limit of the mean relative entropy.

These results extend to the system density matrix. In the time continuous
scaling limit, the system density matrix is a solution of a diffusive Belavkin
equation (\ref{def:belavkin_eq}), as expected. Although not a martingale,
properly modified, it converges to the density matrix predicted by Von Neumann measurement theory.

The article is organized as follow : In section~\ref{sec:QND definition} we define the repeated QND measurement process we study. In section~\ref{sec:apparatus Bayes' law} we establish the link with a classical random process in which the pointer state distribution is repeatedly updated through Bayes' law. In section~\ref{sec:convergence} we prove the convergence of the pointer state distribution under some assumptions and we determine the convergence rate in general cases. In section~\ref{sec:degeneracy} we extend these results to the degenerate case. Finally in section~\ref{sec:continuous time} we study the time continuous scaling limit of our model. Some technical details appearing along the article are postponed to appendices.

\section{QND measurements as stochastic processes}\label{sec:QND definition}

The aim of this section is to describe the relation between repeated non-demolition measurements, positive operator valued measurements (POVM's) and classical stochastic processes.

\subsection{Repeated indirect quantum measurements}

Let us consider a quantum system with initial density matrix $\rho$. Repeated
non-demolition measurements aim at getting indirectly information on the system
(without demolishing it as a projective quantum measurement \`a la Von Neumann
or a direct connection to a macroscopic apparatus might do).

To gain information, we let the system interact with another quantum
system called \textit{the probe}, and then perform a Von Neumann measurement
on the probe. Assume the probe is initially in the pure state $\vert\Psi\rangle\langle\Psi\vert$.
Let $U$ be the unitary operator, acting on the tensor product Hilbert
space $\mathcal{H}_{{\rm sys.}}\otimes\mathcal{H}_{{\rm probe}}$,
coding for the interaction between the system and the probe. After
interaction, the system and the probe are entangled. Their joint state is 
\[
U(\rho_{0}\otimes\vert\Psi\rangle\langle\Psi\vert)U^{\dagger}
\]
 A perfect non-degenerate projective measurement is then performed
on the probe. That is, one is measuring an observable with a non-degenerate
spectrum $i\in\mathcal{I}$. Let $\{\vert i\rangle\}$ be the corresponding
eigenbasis of $\mathcal{H}_{{\rm probe}}$. If the output of the probe
measurement is $i$, the system state is projected into 
\[
\rho'(i)=\frac{1}{\pi(i)}\,{\langle i\vert U\vert\Psi\rangle\rho\langle\Psi\vert U^{\dagger}\vert i\rangle}
\]
because the probe and the system have been entangled.
This projection occurs with probability
\[
\pi(i):={\rm Tr}[\langle i\vert U\vert\Psi\rangle\rho\langle\Psi\vert U^{\dagger}\vert i\rangle].
\]
 We do not have to worry about cases where $\pi(i)=0$, because these
cases, almost surely, never happen.

The process of \textquotedbl{}interaction plus probe measurement\textquotedbl{}
is an example of a positive operator valued measurement (POVM). Let
us define operators $M_{i}$, acting on the system Hilbert space,
by \[
M_{i}:=\langle i\vert U\vert\Psi\rangle.\]
 They satisfy $\sum_{i}M_{i}^{\dagger}M_{i}=\mathbb{I}_{{\rm sys.}}$
as a consequence of the unitarity of $U$. After measurement with
output $i$, the system density matrix $\rho'(i)$ can be written
as \[
\rho'(i)=\frac{1}{\pi(i)}\,{M_{i}\,\rho\, M_{i}^{\dag}},\]
 with $\pi(i)={\rm Tr}[M_{i}\,\rho\, M_{i}^{\dag}]$. This characterizes
a POVM.

\vspace{0.3cm}

Let us now assume that we repeat the process of \textquotedbl{}interaction
plus probe measurement\textquotedbl{} ad libitum. As we shall see
below, even for purely practical reasons, it is useful to keep the
freedom of changing some or all features of the process. For instance,
the experimenter might tune (or let fluctuate randomly, or tune but
leaving a certain amount of randomness or ...) the initial state $\vert\Psi\rangle$
of the probe at each step. Or he/she might tune (or let fluctuate randomly,
or ...) the interaction operator $U$ at each step, for instance
by playing on the time lapse that the probe spends close enough to
the system to interact significantly with it. He/She might even tune (or
...) the type of probe (in particular the dimension of its Hilbert
space) at each step. Finally, he/she might tune (or ...) the non-degenerate probe measurement (equivalently the $\mathcal H_{\rm probe}$ basis made of its eigenvectors $\{\vert i\rangle\}$).

We let $\vert\Psi_{n}\rangle$, $U_{n}$, $\mathcal{I}_{n}$ denote
the initial state, interaction operator and set of possible outcomes
of the ${\rm n}^{\rm th}$ step. Setting $\rho_{0}:=\rho$, $\rho_{1}:=\rho'$
and so on, we get a random recursion equation, namely that for $i\in\mathcal{I}_{n}$:
\begin{eqnarray}
\rho_{n}=\frac{1}{\pi_{n}(i)}\,{M_{i}^{(n)}\rho_{n-1}\,{M_{i}^{(n)}}^{\dagger}}\label{recurrho}\end{eqnarray}
 with probability $\pi_{n}(i)={\rm Tr}[M_{i}^{(n)}\rho_{n-1}{M_{i}^{(n)}}^{\dagger}]$
where $M_{i}^{(n)}=\langle i\vert U_{n}\vert\Psi_{n}\rangle$ (note
that the meaning of the expectation itself, i.e. the Hilbert space
with respect to which it is taken, might depend on $n$, however we
may arrange to choose the $\mathcal{I}_{n}$'s so that $i$ determines
$\langle i\vert$ completely).

It is worth noticing that, under such an evolution, a pure state remains a pure state, that is: $\vert\phi_{n}\rangle={M_{i}^{(n)}\vert\phi_{n-1}\rangle}/{\|M_{i}^{(n)}\vert\phi_{n-1}\rangle\|}$
with probability $\langle\phi_{n-1}\vert{M_{i}^{(n)}}^{\dagger}M_{i}^{(n)}\vert\phi_{n-1}\rangle$.
This case is included in that of density matrices.

\vspace{0.3cm}

Let us now specialize this scheme in such a way
that it preserves a preferred basis of the system Hilbert space. That
is, we assume there exists a fixed basis $\{\vert\alpha\rangle\}$
of $\mathcal{H}_{{\rm sys.}}$ such that all interactions can be decomposed
as 
\begin{eqnarray}
U_{n}:=\sum_{\alpha}\vert\alpha\rangle\langle\alpha\vert\otimes U_{n}(\alpha)\label{Upointer}
\end{eqnarray}
 where the $U_{n}(\alpha)$'s are unitary operators on $\mathcal{H}_{{\rm probe}_{n}}$.
The states $\vert\alpha\rangle$ are called \textit{pointer states}.
The density matrices $\vert\alpha\rangle\langle\alpha\vert$ with
$|\alpha\rangle$ a pointer state are fixed points of the recursion
relation (\ref{recurrho}).

The pointer states have to be eigenstates of the system Hamiltonian
$H_{s}$ for these indirect measurements to be quantum non-demolition
(QND) measurements, since there is a waiting time between two successive indirect measurements during which the quantum system evolves freely.
That is, $H_{s}=\sum_{\alpha}E_{\alpha}\vert\alpha\rangle\langle\alpha\vert$
where $E_{\alpha}$ is the energy of the pointer state $\vert\alpha\rangle$
for the  free system.

After each indirect measurement one gains information on the system
state. Repeating the process (infinitely) many times amounts (as we
shall explain) to perform a measurement of a system observable whose
eigenstates are the pointer states. This observable commutes with the system
free evolution. A system in one of the pointer states remains unchanged
by the successive indirect measurements.

It has been shown in ref.\cite{bauer_bernard} that a system subject
to repeated QND measurements as described above converges toward one
of the pointer states. This convergence was only proved in the case
where the probes, interactions and observables on the probes are all
the same. A non-degeneracy hypothesis was also used. One of the present article
aim is to generalize the convergence statements
without those assumptions.

A word on terminology: we are going to name \textit{partial measurement}
one iteration of \textquotedbl{}interaction plus probe measurement\textquotedbl{}
and \textit{complete measurement} an infinite sequence of successive
partial measurements.

\subsection{A toy model}

We shall illustrate this framework and our results with a simple toy
model inspired by experiments done on quantum electrodynamics in cavities\cite{haroche}.
The present work is actually inspired by these experiments.
There, the system is a monochromatic photon field and the probes are
modeled by two level systems. The observable we aim at measuring is
the photon number. This is a non-demolishing measurement. 

The system-probe interaction is well described by the unitary operator \begin{align*}
U & =\exp[\,{-i(\epsilon\Delta t\,\hat{p}\otimes\mathbb{I}_{2}+\frac{\pi}{4}\hat{p}\otimes\sigma_{3}})\,]\end{align*}
 where $\hat{p}$ is the photon number operator, $\epsilon$ the energy
of a photon and $\Delta t$ the interaction duration.

This interaction amounts to the rotation of the two level system effective spin half
if the cavity happened to be in a photon number operator eigenstate. The probes are assumed to be initially
in a state $\vert\Psi\rangle=e^{-i\theta\sigma_{3}}\vert+_{1}\rangle$
where $\vert+_{1}\rangle$ is an eigenvector of $\sigma_{1}$ corresponding
to the eigenvalue $+1$. The probe observables, which are measured
after the interaction between the system and the probe has taken place, are $O_{\theta'}=e^{-i\theta'\sigma_{3}}\sigma_{2}e^{i\theta'\sigma_{3}}$.
Their eigenvectors are $\vert\pm_{\theta}'\rangle=e^{-i\theta'\sigma_{3}}\vert\pm_{2}\rangle$.

The resulting POVM operators for the process of \textquotedbl{}interaction
plus probe measurement\textquotedbl{} only depend on the difference
between the two angles $\theta$ and $\theta'$. They are $M_{\pm}^{\theta-\theta'}$
with \begin{align*}
M_{\pm}^{\theta-\theta'}= & \langle\pm_{2}\vert e^{i\theta'\sigma_{3}}e^{-i(\epsilon\Delta t\,\hat{p}\otimes\mathbb{I}_{2}+\frac{\pi}{4}\hat{p}\otimes\sigma_{3})}e^{-i\theta\sigma_{3}}\vert+_{1}\rangle\\
= & e^{-i\epsilon\Delta t\,\hat{p}}\langle\pm_{2}\vert e^{i(\theta'-\theta)\sigma_{3}-i\frac{\pi}{4}\hat{p}\otimes\sigma_{3}}\vert+_{1}\rangle\end{align*}
 Using the identity $e^{-i\theta\sigma_{3}}=\cos(\theta)\mathbb{I}_{2}-i\sin(\theta)\sigma_{3}$
and $\sigma_{3}\vert+_{1}\rangle=\vert-_{1}\rangle$, one gets \[
M_{\pm}^{\theta-\theta'}=\frac{1}{\sqrt{2}}\, e^{-i(\epsilon\Delta t\,\hat{p}\pm\frac{\pi}{4})}\big[\cos(\theta-\theta'+\frac{\pi}{4}\hat{p})\pm\sin(\theta-\theta'+\frac{\pi}{4}\hat{p})\big]\]
 One may verify that $M_{+}^{\theta-\theta'\,\dag}M_{+}^{\theta-\theta'}+M_{-}^{\theta-\theta'\,\dag}M_{-}^{\theta-\theta'}=\mathbb{I}$.
Remark that if $\vert p\rangle,p\in\mathbb{N}$ is a fixed photon
number state, then $|\langle p\vert M_{\pm}^{\theta}\vert p\rangle|$
is identical to $|\langle p+4k\vert M_{\pm}^{\theta}\vert p+4k\rangle|$,
with $k\in\mathbb{N}$. This property leads to degeneracies in iterative
QND measurement methods. Two states $\vert p\rangle$ and $\vert p+4k\rangle$
can not be distinguished by this method. These degeneracies are discussed
in section \ref{sec:state_distrib_conv}.

\section{Measurement apparatus and Bayes' law}\label{sec:apparatus Bayes' law}

The aim of this section is to reformulate (part of) of the iterative
QND measurement method in terms of classical probability theory. 
We are interested in the eigenstate probability distribution $q_{n}(\alpha)$, with
\[
q_{n}(\alpha):=\langle\alpha\vert\rho_{n}\vert\alpha\rangle,\]
 and its evolution during the iterative procedure. At each step, the
system density matrix is updated via the relation (\ref{recurrho}),
and as a consequence of the factorization property (\ref{Upointer}),
this distribution is updated through the random recursion relation:
\begin{eqnarray*}
q_{n}(\alpha)=q_{n-1}(\alpha)\frac{|M^{(n)}(i|\alpha)|^{2}}{\sum_{\beta}q_{n-1}(\beta)|M^{(n)}(i|\beta)|^{2}}\end{eqnarray*}
 with probability $\sum_{\beta}q_{n-1}(\beta)|M^{(n)}(i|\beta)|^{2}$.
We have defined $M^{(n)}(i|\alpha):=\langle i\vert U_{n}(\alpha)\vert\Psi_{n}\rangle$.

Since $|M^{(n)}(i|\alpha)|^{2}$ is the probability to get a probe
measurement result $i$ conditioned on the system state being $\vert\alpha\rangle$,
we introduce a (hopefully) suggestive notation \[
p_{n}(i|\alpha):=|M^{(n)}(i|\alpha)|^{2}\]
 We have $\sum_{i}p_{n}(i|\alpha)=1$ since $\sum_{i}M^{(n)}(i|\alpha)^{\dag}M^{(n)}(i|\alpha)=\mathbb{I}$.
The recursion relation on the distribution reads \begin{eqnarray}
q_{n}(\alpha)=q_{n-1}(\alpha)\frac{p_{n}(i|\alpha)}{\sum_{\beta}q_{n-1}(\beta)p_{n}(i|\beta)}\label{recurqn}\end{eqnarray}
 with probability $\pi_{n}(i)=\sum_{\beta}q_{n-1}(\beta)p_{n}(i|\beta)$.
This update rule corresponds to Bayes' law. The study of the eigenstate distribution convergence
is thus a question of classical probability theory.
\vspace{0.3cm}

Let us now put on stage the classical probability theory framework we shall be using.

We imagine building a measurement apparatus which performs a sequence
of partial measurements. As we have stressed above, we allow for a
protocol where the characteristics of partial measurements may vary
at each step. However, we shall assume that these characteristics
are chosen within some finite set $\mathcal{O}$ called the set of
measurement methods. In a quantum setting, one measurement method $o$ is a triplet (probe state $\vert \Psi \rangle$, interaction $U$, probe eigen-basis $\{\vert i\rangle\}$ of direct measurement). Each measurement method $o\in{\mathcal{O}}$
defines a set, called the spectrum of $o$ and denoted by $\spec(o)$,
of possible outcomes. For each $o$ we have a family of probability
measures $p^{o}(\cdot|\alpha)$ on $\spec(o)$ indexed by $\alpha\in{\mathcal{S}}$,
where $\mathcal{S}$ is the index set of pointer states.

As time goes by, the experimenter records the sequence $o_{1},i_{1},o_{2},i_{2},\cdots$
where $o_{1}$ is the first measurement method, $i_{1}$ the outcome
of the first partial measurement, performed using method $o_{1}$,
and so on. So it is natural to take as the space of events the space
$\Omega$ of infinite sequences $(o_{1},i_{1},o_{2},i_{2},\cdots)$
where each $o_{n}$ belongs to $\mathcal{O}$ and each $i_{n}$ to
$\spec(o_{n})$. To be even more formal, set $E:=\cup_{o\in{\mathcal{O}}}\{o\}\times\spec(o)$,
so that $E$ is the set of couples $(o,i)$ with $o\in{\mathcal{O}}$
and $i\in\spec(o)$. Then $\Omega:=E^{{\mathbb{N}}^{*}}$.

For a finite sequence $(o_{1},i_{1},o_{2},\cdots,i_{n},o_{n+1})\in E^{n}\times{\mathcal{O}}$,
$B_{o_{1},i_{1},o_{2},\cdots,i_{n},o_{n+1}}$ is defined as the
subset of $\Omega$ made of all those $\omega$'s whose first $2n+1$
components are $o_{1},i_{1},o_{2},\cdots,i_{n},o_{n+1}$. We define
$B_{o_{1},i_{1},o_{2},\cdots,o_{n},i_{n}}$ analogously. We let ${\mathcal{F}}_{n}$
be the $\sigma$-algebra generated by all the $B_{o_{1},i_{1},o_{2},\cdots,i_{n},o_{n+1}}$.
Note that ${\mathcal{F}}_{0}$ is the $\sigma$-algebra generated
by all the $B_{o_{1}}$, i.e. ${\mathcal{F}}_{0}$ codes for the first measurement method choice. For convenience we define ${\mathcal{F}}_{-1}\equiv\{\emptyset,\Omega\}$.
Then ${\mathbb{F}}:=({\mathcal{F}}_{-1},{\mathcal{F}}_{0},{\mathcal{F}}_{1},\cdots)$
is an increasing sequence of $\sigma$-algebras. We take ${\mathcal{F}}$
to be the smallest $\sigma$-algebra on $\Omega$ containing all the
${\mathcal{F}}_{n}$, making $(\Omega,{\mathcal{F}},{\mathbb{F}})$
a filtered measurable space. We could define another filtration by
taking ${\mathcal{F}}'_{n}$ to be the $\sigma$-algebra generated
by all the $B_{o_{1},i_{1},o_{2},\cdots,o_{n},i_{n}}$. While this
may seem superficially a more natural choice of filtration, we shall
see below that ${\mathbb{F}}$ is slightly more convenient. There
is a natural collection of measurable functions on $(\Omega,{\mathcal{F}})$,
namely the projections : for $\omega=(o_{1},i_{1},o_{2},\cdots,o_{n},i_{n},\cdots)$
we set $O_{n}(\omega)=o_{n}$, $I_{n}(\omega)=i_{n}$. These can be
used to define counting functions that play an important role in the
following. We set $\epsilon_{n}(o,i):={\mathbb{1}}_{O_{n}=o,I_{n}=i}$
and $N_{n}(o,i):=\sum_{1\leq m\leq n}\epsilon_{m}(o,i)$ (with the
usual empty sum convention $N_{0}(o,i):=0)$.

The first task is to put a probability measure on $\Omega$. The next
one will be to define a sequence of random variables on $\Omega$
solving the recursion relation (\ref{recurqn}).

If the measurement methods are given, the distributions of partial
measurements are described by the $p^{o}(\cdot|\alpha)$. So what
remains to be discussed is how the measurement methods are chosen,
and we put the condition that this does not involve precognition.
We suppose that a collection of non-negative numbers $d_{-1}=1$,
$d_{0}(o_{1})$ (for all $o_{1}\in\mathcal{O}$), $\cdots$, $d_{n}(o_{1},i_{1},o_{2},\cdots,i_{n},o_{n+1})$
(for all $o_{1}\in\mathcal{O}$, $i_{1}\in\spec(o_{1})$, $\cdots$,
$o_{n+1}\in\mathcal{O}$) is given in such a way that \begin{equation}
\sum_{o_{n+1}\in\mathcal{O}}d_{n}(o_{1},i_{1},o_{2},\cdots,i_{n},o_{n+1})=d_{n-1}(o_{1},i_{1},o_{2},\cdots,i_{n-1},o_{n}).\label{eq:sum}\end{equation}

It is not difficult to produce such a collection. For instance if
$c_{0}(\cdot)$ is a probability measure on $\mathcal{O}$ and for
each $n\geq1$ $c_{n}(\cdot|o,i)$ is a probability measure on $\mathcal{O}$
indexed by $(o,i)\in E$ then 
\[
d_{n}(o_{1},i_{1},o_{2},\cdots,i_{n},o_{n+1})
\equiv c_{0}(o_{1})c_{1}(o_{2}|o_{1},i_{1})\cdots c_{n}(o_{n+1}|o_{n},i_{n})
\]
does the job. We call this special choice the Markovian feedback
protocol. Some special cases are of interest. If we assume that for
$n\geq1$ $c_{n}(\cdot|o,i)=c_{n}(\cdot)$ does not depend on $o,i$,
we arrive at something we call the random protocol. On the other hand,
if $b_{0}$ is an element of $\mathcal{O}$ and $b_{n}$, $n\geq1$
a family of maps from $E$ to $\mathcal{O}$, then taking $c_{0}(\cdot):=\delta_{\cdot,b_{0}}$
and, for $n\geq1$, $c_{n}(\cdot|o,i):=\delta_{\cdot,b_{n}(o,i)}$
we arrive at the description of an experimenter deciding of the next
measurement method by taking into account the previous
measure outcome.
\vspace{0.3cm}

Given a collection of such non negative numbers $d_{n}(o_1,i_1,\cdots,i_n,o_{n+1})$, and using the Kolmogorov extension theorem, it is easy to see that there
is a unique probability measure ${\mathbb{P}}_{\alpha}$ on $(\Omega,{\mathcal{F}})$
such that 
\[
{\mathbb{P}}_{\alpha}(B_{o_{1},i_{1},o_{2},\cdots,i_{n},o_{n+1}})=p^{o_{1}}(i_{1}|\alpha)\cdots p^{o_{n}}(i_{n}|\alpha)d_{n}(o_{1},i_{1},o_{2},\cdots,i_{n},o_{n+1}).
\]
 Indeed, we see that the mandatory consistency condition is fulfilled
: if the left-hand side is summed over $o_{n+1}$ (using (\ref{eq:sum}))
and then over $i_{n}$ the formula for ${\mathbb{P}}_{\alpha}(B_{o_{1},i_{1},o_{2},\cdots,i_{n-1},o_{n}})$
is recovered, which is needed since $B_{o_{1},i_{1},o_{2},\cdots,i_{n-1},o_{n}}$
is the disjoint union of the $B_{o_{1},i_{1},o_{2},\cdots,i_{n},o_{n+1}}$
over the possible $o_{n+1}$ and $i_{n}$. The normalization condition
${\mathbb{P}}_{\alpha}(\Omega)=1$ and the positivity condition are obvious.

Note that in general, conditional on the sequence of measurement methods
$o_{1},\cdots,o_{n}$, one has ${\mathbb{P}}_{\alpha}(i_{1},\cdots,i_{n}|o_{1},\cdots,o_{n})\neq p^{o_{1}}(i_{1}|\alpha)\cdots p^{o_{n}}(i_{n}|\alpha)$.
This is due to the feedback. For the cases when the $d_{n}$'s do not depend on the outcomes, in particular for the independent random protocol, equality is recovered.

We define also,
\begin{equation} \label{eq:defpp}
{\mathbb{P}}\equiv\sum_{\alpha\in{\mathcal{S}}}q_{0}(\alpha){\mathbb{P}}_{\alpha}.
\end{equation}
We use ${\mathbb{E}}_{\alpha}$ and ${\mathbb{E}}$ to denote expectations
with respect to ${\mathbb{P}}_{\alpha}$ and ${\mathbb{P}}$ respectively.

A simple computation shows that, for each $\alpha$, the conditional
probability \[
{\mathbb{P}}_{\alpha}(O_{n+1}=o_{n+1}|O_{1},I_{1},\cdots,O_{n},I_{n})=\frac{d_{n}(O_{1},I_{1},O_{2},\cdots,I_{n},o_{n+1})}{d_{n-1}(O_{1},I_{1},0_{2},\cdots,I_{n-1},O_{n})}\]
 whenever the denominator in nonzero. The same formula holds for ${\mathbb{P}}$.
The right-hand side is simply $c_{n}(o_{n+1}|O_{n},I_{n})$ for the
Markovian feedback protocol. So indeed, the functions $d_{-1},d_{0},d_{1},\cdots$
embody the probabilistic description of the choice of measurement
methods.

These definitions may seem arbitrary at that point, but now we can
make contact with the initial problem. Define a sequence of random variables
$Q_{n}(\alpha)$ by the initial condition $Q_0(\alpha)=q_0(\alpha)$ and the
recursion relation
\[
Q_{n}(\alpha)=\frac{Q_{n-1}(\alpha)p^{O_{n}}(I_{n}|\alpha)}{\sum_{\beta\in{\mathcal{S}}}Q_{n-1}(\beta)p^{O_{n}}(I_{n}|\beta)}.\]
To show that the recursion relation (\ref{recurqn}) is verified, we need to show
that the transition probabilities are correct. A simple way to do that is to
solve this random recursion relation. 
For $\omega\in B_{o_{1},i_{1},o_{2},\cdots,i_{n},o_{n+1}}$ one checks that 
\begin{equation}
Q_{n}(\alpha,\omega):=\frac{q_{0}(\alpha)p^{o_{1}}(i_{1}|\alpha)\cdots
  p^{o_{n}}(i_{n}|\alpha)}{\sum_{\beta\in{\mathcal{S}}}q_{0}(\beta)p^{o_{1}}(i_{1}|\beta)\cdots p^{o_{n}}(i_{n}|\beta)}\label{eq:explicit_def_Qn}
\end{equation}
whenever ${\mathbb{P}}(B_{o_{1},i_{1},o_{2},\cdots,i_{n},o_{n+1}})\neq0$.
Note that this condition ensures that the denominator of $Q_{n}(\alpha,\omega)$
is nonzero. We observe that whenever defined, $Q_{n}(\alpha,\omega)\geq0$
and $\sum_{\alpha}Q_{n}(\alpha,\omega)=1$. If ${\mathbb{P}}(B_{o_{1},i_{1},o_{2},\cdots,i_{n},o_{n+1}})=0$
the value of $Q_{n}(\alpha,\omega)$ is mostly immaterial from a probabilistic
viewpoint, because in any case the full sequence $Q_{n}(\alpha,\omega)$
is well-defined on a set $\tilde{\Omega}$ of ${\mathbb{P}}$-measure
$1$ (note that the collection of all $B_{o_{1},i_{1},o_{2},\cdots,i_{n},o_{n+1}}$
is countable, so the collection of those with ${\mathbb{P}}$-measure
$0$ is countable as well, or empty).

Since there is no dependence on $o_{n+1}$ on the right-hand
side we observe that for $\omega\in\tilde{\Omega}$,
and conditional on ${\mathcal F}_{n-1}$,
\[
Q_{n}(\alpha)=\frac{Q_{n-1}(\alpha)p^{o_{n}}(i_{n}|\alpha)}{\sum_{\beta\in{\mathcal{S}}}Q_{n-1}(\beta)p^{o_{n}}(i_{n}|\beta)}\]
 with probability \[
\sum_{o_{n+1}\in{\mathcal{O}}}\frac{{\mathbb{P}}(B_{o_{1},i_{1},o_{2},\cdots,i_{n},o_{n+1}})}{{\mathbb{P}}(B_{o_{1},i_{1},o_{2},\cdots,i_{n-1},o_{n}})}=\frac{{\mathbb{P}}(B_{o_{1},i_{1},o_{2},\cdots,i_{n}})}{{\mathbb{P}}(B_{o_{1},i_{1},o_{2},\cdots,i_{n-1},o_{n}})}=\sum_{\beta\in{\mathcal{S}}}Q_{n-1}(\beta)p^{o_{n}}(i_{n}|\beta).\]
 So the recursion relation (\ref{recurqn}) is recovered with the
identifications $q_{n}(\alpha)\rightarrow Q_{n}(\alpha)$ and $p_{n}\rightarrow p^{o_{n}}$.
To summarize, we have accomplished our goal : find a probability space
on which (\ref{recurqn}) has a solution, which we have even written
explicitly.

In the following sections we study the convergence of $Q_{n}(.)$
and its dependence with respect to the initial pointer state distribution.
On our way, we shall understand the probabilistic meaning
of the recursion relation (\ref{recurqn}).

\section{Convergence}\label{sec:convergence}

In \cite{bauer_bernard} the convergence of $q_{n}(.)$ has been shown
under the hypothesis that only one partial measurement method $o$
is used. The properties of the limit were elucidated under the further
assumption that for every couple of pointer states $(\alpha,\beta)$
there exists at least one partial measurement result $i$ such that
$p^{o}(i|\alpha)\ne p^{o}(i|\beta)$. This last assumption can be
understood as a non-degeneracy hypothesis because two different pointer states
$\alpha,\beta$ do not induce identical partial measurement results
distribution $p^{o}(i|\alpha)$. Our aim is to generalize the convergence
of $Q_{n}(.)$ while weakening the hypotheses made in \cite{bauer_bernard}.
We discuss the convergence when different partial measurement methods
are used. We focus on the influence of this extension on the rate
of convergence. The degenerate case will be studied in section \ref{sec:state_distrib_conv}.
In the case of one measurement method, a convergence result similar to that of \cite{bauer_bernard} has been obtained by H. Amini, P. Rouchon and M. Mirrahimi through sub-martingale convergence in \cite{H. Amini}.

\subsection{Convergence with different partial measurement methods}

\label{sec:conv} The extension of the convergence result of \cite{bauer_bernard}
to cases with different measurement methods is straightforward.
From the fact that, conditional on $\mathcal F_{n-1}$, 
\[
Q_{n}(\alpha)=\frac{Q_{n-1}(\alpha)p^{o_{n}}(i_{n}|\alpha)}{\sum_{\beta\in{\mathcal{S}}}Q_{n-1}(\beta)p^{o_{n}}(i_{n}|\beta)}
\]
with probability $\sum_{\beta\in{\mathcal{S}}}Q_{n-1}(\beta)p^{o_{n}}(i_{n}|\beta)$,
the average of $Q_{n}(\alpha)$, again conditioned on $\mathcal F_{n-1}$, is 
\[
\sum_{i_{n}\in\spec{o_{n}}}Q_{n}(\alpha)\sum_{\beta\in{\mathcal{S}}}Q_{n-1}(\beta)p^{o_{n}}(i_{n}|\beta)=Q_{n-1}(\alpha).\]
 So $Q_{n-1}(\alpha)$ is conserved in average. Though the computation involved
to prove it is essentially the same, a mathematically cleaner statement
is that \[
{\mathbb{E}}(Q_{n}(\alpha)\vert{\mathcal{F}}_{n-1})=Q_{n-1}(\alpha),\]
 i.e. each $Q_{n}(\alpha)$ is an $\mathbb{F}$-martingale.
\vspace{0.3cm}

In fact, $Q_{n}(\alpha)$ has a deeper probabilistic meaning, which
makes the martingale property obvious.

For a while, forget the previous definition of $Q_{n}(\alpha)$.
Observe that, under the assumption that $q_{0}(\alpha)>0$ for every
$\alpha\in{\mathcal{S}}$, any set of ${\mathbb{P}}$-measure $0$
has also ${\mathbb{P}}_{\alpha}$-measure $0$. Then the Radon-Nikodym
theorem states that for each $\alpha\in{\mathcal{S}}$, there is a
${\mathbb{P}}$-integrable non-negative random variable $Q(\alpha)$
on $(\Omega,{\mathcal{F}})$ such that 
\[q_{0}(\alpha){\mathbb{E}}_{\alpha}(X)={\mathbb{E}}(Q(\alpha)X)\]
for every ${\mathbb{P}}_{\alpha}$-integrable random variable $X$
on $(\Omega,{\mathcal{F}})$. The random variable $Q(\alpha)$ is
a Radon-Nikodym derivative of $q_{0}(\alpha){\mathbb{P}}_{\alpha}$
with respect to ${\mathbb{P}}$. It is obvious that two Radon-Nikodym
derivatives can differ only on a set of ${\mathbb{P}}$-measure $0$:
in that sense the Radon-Nikodym derivative is unique if it exists.
We have also that, ${\mathbb{P}}$-almost surely, $\sum_{\alpha}Q(\alpha)=1$,
so that, ${\mathbb{P}}$-almost surely, each $Q(\alpha)\leq1$. This
existence theorem is a bit abstract but if one replaces ${\mathcal{F}}$
by ${\mathcal{F}}_{n}$ one can get a concrete formula. The same argument
ensures the existence of a ${\mathbb{P}}$-integrable non-negative
random variable $Q_{n}(\alpha)$ on $(\Omega,{\mathcal{F}}_{n})$
such that $q_{0}(\alpha){\mathbb{E}}_{\alpha}(X)={\mathbb{E}}(Q_{n}(\alpha)X)$
for every ${\mathbb{P}}_{\alpha}$-integrable random variable $X$
on $(\Omega,{\mathcal{F}}_{n})$. As ${\mathcal{F}}_{n}$ is finite,
it suffices to let $X$ run over the indicator functions for the $B_{o_{1},i_{1},o_{2},\cdots,i_{n},o_{n+1}}$.
This implies that \[
Q_{n}(\alpha,\omega)=q_{0}(\alpha)\frac{{\mathbb{P}}_{\alpha}(B_{o_{1},i_{1},o_{2},\cdots,i_{n},o_{n+1}})}{{\mathbb{P}}(B_{o_{1},i_{1},o_{2},\cdots,i_{n},o_{n+1}})}\]
 for every $\omega\in B_{o_{1},i_{1},o_{2},\cdots,i_{n},o_{n+1}}$
(such that the denominator is nonzero, else the value of $Q_{n}(\alpha,\omega)$
is immaterial). Explicitly one finds \[
Q_{n}(\alpha,\omega)=\frac{q_{0}(\alpha)p^{o_{1}}(i_{1}|\alpha)\cdots p^{o_{n}}(i_{n}|\alpha)}{\sum_{\beta\in{\mathcal{S}}}q_{0}(\beta)p^{o_{1}}(i_{1}|\beta)\cdots p^{o_{n}}(i_{n}|\beta)}\]
 on $B_{o_{1},i_{1},o_{2},\cdots,i_{n},o_{n+1}}$.

This is exactly our previous definition of $Q_{n}(\alpha)$, which
is probabilistically a Radon-Nikodym derivative. This makes the martingale
property obvious without any computation, just by the definition and
general properties of conditional expectations. In fact, $Q_{n}(\alpha)$
is a closed martingale: 
\[Q_{n}(\alpha)={\mathbb{E}}(Q(\alpha)|{\mathcal{F}}_{n}).\]
As $Q_{n}(\alpha)$ is also bounded, the martingale convergence theorem \cite{bernt_oksendal}
ensures that $Q_{n}(\alpha)\rightarrow Q(\alpha)$ $\mathbb{P}$-almost
surely and in ${\mathbb{L}}^{1}(\Omega,{\mathcal{F}},\mathbb{P})$.
Since by assumption $\mathcal{S}$ is a finite set, all $Q_{n}(\alpha)$'s
converge simultaneously $\mathbb{P}$-almost surely.

\vspace{0.3cm}

Let us now observe that the following two statements are equivalent:

--- the measures ${\mathbb{P}}_{\alpha}$ are all mutually singular,

--- there is a collection $(\Omega_{\alpha})_{\alpha\in{\mathcal{S}}}$ of
disjoint measurable subsets of $\Omega$ such that, $\mathbb{P}$-almost surely
${\mathbb{1}}_{\Omega_{\alpha}}=Q(\alpha)$.

\vspace{0.1cm}

The proof is simple. The statement that the measures ${\mathbb{P}}_{\alpha}$ are
all mutually singular is equivalent to the existence of a collection
$(\Omega_{\alpha})_{\alpha\in{\mathcal{S}}}$ of disjoint measurable subsets of
$\Omega$ such that ${\mathbb{P}}_{\beta}(\Omega_{\alpha})=\delta_{\alpha,\beta}$
for all $\alpha,\beta\in{\mathcal{S}}$.  From the defining property of
Radon-Nikodym derivatives, ${\mathbb{1}}_{\Omega_{\alpha}}Q(\alpha)$ is also a
Radon-Nikodym derivative of $q_{0}(\alpha){\mathbb{P}}_{\alpha}$ with respect to
${\mathbb{P}}$, and
${\mathbb{E}}(Q(\alpha){\mathbb{1}}_{\Omega_{\beta}})=q_{0}(\alpha)
\delta_{\alpha,\beta}$ which by positivity implies that, for $\alpha\neq\beta$,
$Q(\alpha){\mathbb{1}}_{\Omega_{\beta}}=0$ except maybe on a set of
${\mathbb{P}}$-measure $0$. Hence $\mathbb{P}$-almost surely
${\mathbb{1}}_{\Omega_{\alpha}}Q(\beta)=Q(\beta)\delta_{\alpha,\beta}$.  Summing
over $\beta$ gives ${\mathbb{1}}_{\Omega_{\alpha}}=Q(\alpha)$
$\mathbb{P}$-almost surely. The converse is also true: if there is a collection
$(\Omega_{\alpha})_{\alpha\in{\mathcal{S}}}$ of disjoint measurable subsets of
$\Omega$ such that $\mathbb{P}$-almost surely
${\mathbb{1}}_{\Omega_{\alpha}}=Q(\alpha)$, then the measures
${\mathbb{P}}_{\alpha}$ are all mutually singular and concentrated on the
$\Omega_{\alpha}$'s.

A striking consequence is that, if the measures ${\mathbb{P}}_{\alpha}$ are all
mutually singular, for each $\omega$ in a set of ${\mathbb{P}}$-measure $1$,
$Q_{n}(\alpha)$ converges to either $0$ or $1$, and it converges to $1$ with
probability ${\mathbb{P}}(\Omega_{\alpha})=q_{0}(\alpha)$. 

Hence when the measures ${\mathbb{P}}_{\alpha}$ are all mutually
singular \textit{there is a full experimental equivalence between
an infinite sequence of partial measurements and a direct projective
measurement on the system.} We further study this equivalence in section
\ref{sec:early_proj}.

\vspace{0.3cm}

We shall now give a criterion, that the experimenter
may enforce on the protocol, ensuring that the measures ${\mathbb{P}}_{\alpha}$
are all mutually singular. This involves a non-degeneracy hypothesis, similar but weaker than that made in \cite{bauer_bernard}.

\vspace{0.1cm}

We say that $o \in \mathcal{O}$ is recurrent in
$\omega\in\Omega$ if $O_{n}(\omega)=o$ for infinitely many $n$'s.
Our (sufficient) criterion for all the measures ${\mathbb{P}}_{\alpha}$
to be mutually singular is:  

There is a subset ${\mathcal O}_{s}$ of ${\mathcal O}$ such that

-- Each $o\in {\mathcal O}_{s}$ is recurrent with probability $1$ under each
${\mathbb{P}}_{\alpha}$

-- For every $\alpha,\beta\in{\mathcal{S}}$, $\alpha\neq\beta$ there
is some $o \in {\mathcal O}_{s}$ and $i \in \spec (o)$ such that $p^{o}(i|\alpha)\neq p^{o}(i|\beta)$. 

This condition says that with probability one, infinitely many partial measurements 
that distinguish between any two states of the system will occur.  

\vspace{0.1cm}

Consider the event $A_{o}$ made of the $\omega$'s such that $Q_n(\alpha)$
converges for each $\alpha$ and $o$ is recurrent. Note that by our assumptions
${\mathbb P}(A_{o})=1$. We show that for any $i\in \spec (o)$
\[
Q(\alpha)\sum_{\gamma}Q(\gamma)p^{o}(i|\gamma)=Q(\alpha)p^{o}(i|\alpha)
\]
on $A_{o}$. There are two cases to consider. Either
$(O_{n}(\omega),I_{n}(\omega))=(o,i)$ for infinitely many $n$'s: then the
announced relation follows by taking the limit of the basic recursion relation
along a subsequence. Or $(O_{n}(\omega),I_{n}(\omega))=(o,i)$ for only finitely
many $n$'s: then, as shown in Appendix \ref{sec:details},
$\sum_{\gamma}Q(\gamma)p^{o}(i|\gamma)=0$ so that in particular
$Q(\alpha)p^{o}(i|\alpha)=0$ and the announced relation still holds.

This implies that
\begin{equation}
\forall\,\alpha,\beta\in{\mathcal{S}},\;
Q(\alpha)Q(\beta)(p^{o}(i|\alpha)-p^{o}(i|\beta))=0\text{ on
}A_{o} \text{ for every } i \in \spec (o).\label{eq:separ}
\end{equation}
Then, by (\ref{eq:separ}),
$\forall\,\alpha,\beta\in{\mathcal{S}},\;\alpha\neq\beta$ one has
$Q(\alpha)Q(\beta)=0$ on $A_{{\mathcal O}_{s}}:=\cap_{o \in {\mathcal
    O}_{s}}A_{o}$, which has ${\mathbb{P}}$-measure $1$. As the sum of the
$Q(\alpha)$'s is $1$, this means that on $A_{{\mathcal O}_{s}}$ one has
$Q(\alpha)=\delta_{\alpha,\Upsilon}$ where $\Upsilon(\omega)$ is some
$\omega$-dependent element of $\mathcal{S}$.  So there is a family of disjoint
subsets $\Omega_{\alpha}$ of $A_{{\mathcal O}_{s}}$ such that
$\cup_{\alpha}\Omega_{\alpha}$ has ${\mathbb{P}}$-measure $1$, and
$Q(\alpha)={\mathbb{1}}_{\Omega_{\alpha}}$ except maybe on a set of
${\mathbb{P}}$-measure $0$.

\vspace{0.3cm}

We shall give two examples.

For the first one, the task to ensure that enough measurement methods $o$ are
recurrent is left to the experimenter.

The second example is the Markovian feedback protocol. For this protocol, under
${\mathbb{P}}_{\alpha}$, the process $(o_{n},i_{n})$ is Markovian with
transition kernel $K_{\alpha,n}(o,i;o',i')=p^{o'}(i'|\alpha)c_{n}(o'|o,i)$, with
initial distribution $p^{o}(i|\alpha)c_{0}(o)$. Recurrence questions are well
under control at least when the kernels do not depend on $n$. So we assume that
$c_{n}(o'|o,i)=c(o'|o,i)$ is time independent%
\footnote{Were $c_{n}$ periodic in $n$ we could look at a Markov chain with a
  larger state space to reduce to that case.%
} and set $K_{\alpha}(o,i;o',i')=p^{o'}(i'|\alpha)c(o'|o,i)$. The product
structure of $K_{\alpha}$ leads to introduce the reduced transition kernel
$K_{\alpha}^{red}(o;o'):=\sum_{i\in\spec(o)}p^{o}(i|\alpha)c(o'|o,i)$. One
can rely on classical Markov chain computations to make sure that the measures
${\mathbb{P}}_{\alpha}$ are all mutually singular.  Assuming that the reduced
Markov chain $K_{\alpha}^{red}$ is irreducible and aperiodic, it admits a unique
invariant probability $\mu_{\alpha}^{red}$ on $\mathcal{O}$, which is strictly
positive. Then all partial measurement methods with $\mu_{\alpha}^{red}(o)>0$ will be recurrent on a set of $\mathbb P_\alpha$-measure 1. Moreover the full Markov chain has a unique invariant probability
$\mu_{\alpha}(o,i)=p^{o}(i|\alpha)\mu_{\alpha}^{red}(o)$. Then the strong law of large
numbers for Markov chains states that $N_{n}(o,i)$, the number of occurrence of
$(o,i)$ up to the ${\rm n}^{\rm th}$ experiment, satisfies \[ N_{n}(o,i)\sim
n\mu_{\alpha}(o,i)\text{ for large }n\] on a set of
${\mathbb{P}}_{\alpha}$-measure $1$. This ergodicity result will be put in use
in section \ref{sec:conv_rate}.

\vspace{0.3cm}

To summarize, we have proved that if there are enough ${\mathbb{P}}_{\alpha}$
recurrent partial measurement methods then the measures ${\mathbb{P}}_{\alpha}$
are all mutually singular, so that there is a full experimental equivalence
between an infinite sequence of partial measurements and a direct projective
measurement on the system.

\subsection{Conditioning or projecting}
\label{sec:early_proj}

In the previous section we pointed out the connection between complete
measurements and direct projective measurements. This holds whenever
the measures ${\mathbb{P}}_{\alpha}$ are mutually singular.

Under this hypothesis, we show in this section that we can solve the
random recursion relation (\ref{recurqn}) on a space where the final
pointer state is determined before the measurement process starts:
for the class of experiments we are dealing with, it is consistent
to assess that the total measurement outcome can be decided
in advance and by a classical probabilistic choice. 

\vspace{0.3cm}

We assume that the measures ${\mathbb{P}}_{\alpha}$ are mutually
singular, and, as usual, that all $q_{0}(\alpha)$ are $>0$. To avoid
clumsy statements, we remove from $\Omega$ the (${\mathbb{P}}$-negligible)
set of events for which either $Q_{n}(\alpha)$ is not defined for
all $n$'s, or $Q_{n}(\alpha)$ does not converge to $0$ or $1$.
So we assume that the sets $\Omega_{\alpha}$ form a partition of
$\Omega$, and the random variable $\Upsilon$ defined by $\lim_{n\rightarrow\infty}Q_{n}(\alpha,\omega)={\mathbb{1}}_{\omega\in\Omega_{\alpha}}=\delta_{\alpha,\Upsilon(\omega)}$
is defined everywhere on $\Omega$. We let $\mathcal{A}$ be the smallest
$\sigma$-algebra making any $\Omega_{\alpha}$ measurable. We claim
that if $X\in{\mathbb{L}}^{1}(\Omega,{\mathcal{F}},\mathbb{P})$ is
any integrable random variable, \[
{\mathbb{E}}(X|{\mathcal{A}})={\mathbb{E}}_{\Upsilon}(X).\]
 This can be rephrased as : conditioning $\mathbb{P}$ on the limit
of $Q_{n}(\alpha)$ being $1$ leads to ${\mathbb{P}}_{\alpha}$.
This is essentially obvious from the Radon-Nikodym viewpoint and the
fact that the measures ${\mathbb{P}}_{\alpha}$ are mutually singular.
But a direct computation is easy. The fact that $\mathcal{S}$ is finite
(countable would do the job as well) has two consequences. First,
any ${\mathcal{A}}$-measurable random variable $Y$ can be written
as a linear combination $Y=\sum_{\alpha\in{\mathcal{S}}}y_{\alpha}{\mathbb{1}}_{\Omega_{\alpha}}$.
Second, to test that ${\mathbb{E}}(X|{\mathcal{A}})=Y$ it suffices
to check that ${\mathbb{E}}(X{\mathbb{1}}_{\Omega_{\alpha}})={\mathbb{E}}(Y{\mathbb{1}}_{\Omega_{\alpha}})$
for every $\alpha$. Now, by definition, ${\mathbb{E}}(X{\mathbb{1}}_{\Omega_{\alpha}})=q_{0}(\alpha){\mathbb{E}}_{\alpha}(X)$,
whereas ${\mathbb{E}}(Y{\mathbb{1}}_{\Omega_{\alpha}})=q_{0}(\alpha)y_{\alpha}$,
so, if ${\mathbb{E}}(X|{\mathcal{A}})=Y$, $y_{\alpha}={\mathbb{E}}_{\alpha}(X)$,
i.e. $Y=\sum_{\alpha\in{\mathcal{S}}}{\mathbb{E}}_{\alpha}(X){\mathbb{1}}_{\Omega_{\alpha}}={\mathbb{E}}_{\Upsilon}(X)$.
Hence ${\mathbb{E}}(X|{\mathcal{A}})={\mathbb{E}}_{\Upsilon}(X)$
as announced.

This proves the equivalence between projecting first on a given state
$\alpha$ and conditioning on the limit state being $\alpha$.

\subsection{Convergence rates and trial distribution independence}
\label{sec:conv_rate}

Experimentally, the initial distribution $q_{0}(\cdot)$ may not be known. One would then use the sequence of partial measurements to gain information and reconstruct it from these measurements. This
may be done using Bayes' law starting from a trial distribution $\hat{q}_{0}(\cdot)$
(supposed to be nowhere vanishing) and recursively improving it using
the relation 
\[
\hat{q}_{n}(\alpha)=\hat{q}_{n-1}(\alpha)
\frac{p^{o_{n}}(i_{n}|\alpha)}{\sum_{\beta}\hat{q}_{n-1}(\beta)p^{o_{n}}(i_{n}|\beta)}
\]
if the outcome is $i$, which happens with probability
$\sum_{\beta}q_{n-1}(\beta)p^{o_{n}}(i_{n}|\beta)$.  The difference with
eq.(\ref{recurqn}) is that the recursion involves $\hat{q}_{n}(\cdot)$ and not
$q_{n}(\cdot)$. However, the probability is the one given by the $q_{n}(\cdot)$.
If the initial trial distribution $\hat{q}_{0}(\cdot)$ coincides with the
initial system distribution $q_{0}(\cdot)$, then
$\hat{q}_{n}(\cdot)=q_{n}(\cdot)$ for all $n$. Both $\hat{q}_{n}(\cdot)$ and
$q_{n}(\cdot)$ are realization dependent. We shall define the random process
$\hat{Q}_{n}(\cdot)$ as $Q_{n}(\cdot)$ in (\ref{eq:explicit_def_Qn}) but with a
different initial distribution
\[
\hat{Q}_{n}(\alpha):=\hat{q}_{0}(\alpha)
\frac{p^{o_{1}}(i_{1}|\alpha)\cdots p^{o_{n}}(i_{n}|\alpha)}{\sum_{\beta}\hat{q}_{0}(\beta)p^{o_{1}}(i_{1}|\beta)\cdots p^{o_{n}}(i_{n}|\beta)}
\]
 The probability law still depends on the true initial
distribution. Notice that $\hat{Q}_{n}(\cdot)$ is not an $\mathbb F$-martingale under this law,
contrary to $Q_{n}(\cdot)$. As we shall show, they nevertheless have
identical limit, that is: $\lim_{n\to\infty}\hat{Q}_{n}(\alpha)$
exists and is equal to $\mathbb{1}_{\Omega_{\alpha}}$ with $\mathbb
P$-probability 1.

Moreover, if a time independent Markovian feedback protocol is used, the
convergence of the state probability distribution is exponential.  Its
convergence rate is the mean relative entropy of the partial measurement result
distribution conditioned on the system be in the state $\Upsilon$ with respect
to the one conditioned on the system be in the state $\alpha$. This means that
for $n$ large enough,
\begin{eqnarray}
\hat{Q}_{n}(\alpha)\simeq e^{-n\overline{S}(\Upsilon|\alpha)},\quad{\rm for}\ \alpha\ne\Upsilon\label{qdecroit}
\end{eqnarray}
 with 
 \begin{eqnarray}
\overline{S}(\beta|\alpha):=\sum_{o\in\mathcal{O}}\mu_{\Upsilon}^{red}(o)
\sum_{i\in\spec(o)}p^{o}(i|\beta)\ln\left[\frac{p^{o}(i|\beta)}{p^{o}(i|\alpha)}\right]\label{meanS}
\end{eqnarray}
Here, all the $p^{o}(i|\alpha)$ are assumed to be strictly positive, thus any
$(o,i)$ with $o \in \mathcal O_{s}$ is recurrent. In the case of time
independent random protocols, the rate is the same with
$\mu_{\Upsilon}^{red}(\cdot)$ replaced by $c(\cdot)$ the distribution of
measurement methods. This coincides with the result of \cite{bauer_bernard} if
$\mathcal{O}$ contains only one partial measurement method.  \vspace{0.3cm}

The independence of the limiting distribution with respect to the initial trial
distribution is obtained whenever one starts with a trial distribution such that
$\hat{q}_0(\alpha)>0$ wherever $q_0(\alpha)>0$. This happens for example if we
start with $\hat{q}_0(\alpha)>0$ for any $\alpha \in \mathcal S$.

To see this, we analyse the behavior of $\hat{Q}_{n}$ under the probability
measure $\hat{\mathbb P}:=\sum_{\alpha \in \mathcal S} \hat{q}_0(\alpha) \mathbb
P_\alpha$, which can be seen as a trial probability measure on $\Omega$ as it
corresponds to a system initialy in the trial state.  Under $\hat{\mathbb P}$,
$\hat{Q}_{n}(\alpha)$ is a martingale, so by the above arguments, it converges
$\hat{\mathbb P}$ almost surely to $\mathbb{1}_{\Omega_{\alpha}}$. As by
hypothesis $\hat{q}_0(\alpha)>0$ whenever $q_0(\alpha)>0$ (which can be
rephrased as: $\mathbb P$ is absolutely continuous with respect to $\hat{\mathbb
  P}$), a subset of $\Omega$ of $\hat{\mathbb P}$ probability $1$ has also
$\mathbb P$ probability $1$. So $\lim_{n\to\infty}
\hat{Q}_{n}(\alpha)=\mathbb{1}_{\Omega_{\alpha}}=\lim_{n\to\infty} Q_{n}(\alpha)$ with
$\mathbb P$ probability $1$.

What is less direct is the determination of the convergence rate. This requires controlling the behavior of the counting processes $N_{n}(o,i)$.  As recalled at the end of section \ref{sec:conv}, $N_{n}(o,i)/n \rightarrow\mu_{\alpha}(o,i)$ on a set\footnote{For
  the record, the set $\{N_{n}(o,i)/n \rightarrow
\mu_{\alpha}(o,i)\}$ is measurable, as it can be written
$\bigcap_{m\in\mathbb{N}^*}\bigcup_{n_{0}\in\mathbb{N}}\bigcap_{n>n_{0}}\left\{
  \omega\in\Omega, \left|N_{n}(o,i;\omega)/n-\mu_{\alpha}(o,i)
  \right|<1/m\right\}$.} of ${\mathbb{P}}_{\alpha}$-measure $1$. 
We want to infer that $\mathbb{P}$-almost surely,
\[
\lim_{n\to\infty}N_{n}(o,i)/n=\mu_{\Upsilon}(o,i).
\]
To prove it, we set ${\mathcal
  L}(o,i):=\{\lim_{n\to\infty}N_{n}(o,i)/n=\mu_{\Upsilon}(o,i)\}$ and write
\[
{\mathbb P}({\mathcal L}(o,i)) =  \sum_{\alpha} q_0(\alpha) 
{\mathbb P}({\mathcal L}(o,i) | \Omega_{\alpha})= \sum_{\alpha} q_0(\alpha)=1,
\]
where we used 
\begin{align*}
{\mathbb P}({\mathcal L}(o,i)|\Omega_{\alpha})&= {\mathbb P}(\lim_{n\to\infty}N_{n}(o,i)/n=\mu_{\alpha}(o,i)|\Omega_{\alpha})\\
    &={\mathbb P}_\alpha(\lim_{n\to\infty}N_{n}(o,i)/n=\mu_{\alpha}(o,i))=1.
\end{align*}

Observe now that $\hat{Q}_{n}(\alpha)$ can be expressed as a function of the counting
processes:
 \[
 \hat{Q}_{n}(\alpha)=\frac{\hat{q}_{0}(\alpha)\prod_{(o,i)\in
     E}p^{o}(i|\alpha)^{N_{n}(o,i)}}{\sum_{\beta}\hat{q}_{0}(\beta)\prod_{(o,i)\in
     E}p^{o}(i|\beta)^{N_{n}(o,i)}}
\] 
Under the hypothesis that all $p^o(i|\alpha)$'s are >0, we have that $\hat{Q}_n(\alpha)>0$ for every $n$, and the logarithm of the ratio between
$\hat{Q}_{n}(\alpha)$ and $\hat{Q}_{n}(\beta)$ is well defined. Using the
previous result, we have
\begin{align*}
\lim_{n\to\infty}\frac{1}{n}\ln\left[\frac{\hat{Q}_{n}(\beta)}  {\hat{Q}_{n}(\alpha)}\right]&=\sum_{(o,i)\in E}\mu_{\Upsilon} (o,i)\ln\left[\frac{p^{o}(i|\beta)}{p^{o}(i|\alpha)}\right]\\
    &=\sum_{(o,i)\in E}\mu^{red}_{\Upsilon} (o)p^o(i|\Upsilon)\left(\ln\left[\frac{p^{o}(i|\beta)}{p^{o}(i|\Upsilon)}\right] + \ln\left[\frac{p^{o}(i|\Upsilon)}{p^{o}(i|\alpha)}\right]\right)
\end{align*}
Then, for a large enough $n$, 
\[
\frac{\hat{Q}_{n}(\beta)}{\hat{Q}_{n}(\alpha)} \simeq
e^{-n\overline{S}(\Upsilon|\beta)}e^{n\overline{S}(\Upsilon|\alpha)}
\] 
with $\overline{S}(\Upsilon|\alpha)$ the mean relative entropy,
$\overline{S}(\Upsilon|\alpha)=\sum_{o\in\mathcal{O}}\mu_{\Upsilon}^{red}(o)
S^{o}(\Upsilon|\alpha)$ where
\[
S^{o}(\Upsilon|\alpha)=\sum_{i\in\spec(o)}p^{o}(i|\Upsilon)
(\ln[p^{o}(i|\Upsilon)]-\ln[p^{o}(i|\alpha)])
\]
The relative entropy is always non negative, subsequently,
the mean relative entropy is non negative too. Moreover, the mean
relative entropy is null if and only if $\Upsilon=\beta$ (all relative
entropies null).

Using this property and $\sum_{\beta}\hat{Q}_{n}(\beta)=1$, we obtain for $\alpha\ne\Upsilon$, 
\[
{\hat{Q}_{n}(\alpha)}^{-1}=\sum_{\beta}\frac{\hat{Q}_{n}(\beta)}{\hat{Q}_{n}(\alpha)}\simeq
e^{n\overline{S}(\Upsilon|\alpha)}(1+\sum_{\beta\ne\Upsilon}
e^{-n\overline{S}(\Upsilon|\beta)})
\]
Then, to leading exponential order
\[
 \hat{Q}_n(\alpha)\simeq e^{-n\overline{S}(\Upsilon|\alpha)}
\]
Hence, for $n$ large enough, we proved that
\[
\hat{Q}_{n}(\alpha)\simeq\left\{ \begin{array}{ll}
1 & \mbox{if }\alpha=\Upsilon\\
{\rm const.}e^{-n\overline{S}(\Upsilon|\alpha)} & \mbox{else}\end{array}\right.
\]
The limit distribution does not depend on
the trial initial distribution but only on the complete measurement
realization. The probability to have
$\hat{Q}_{\infty}(\alpha)=\delta_{\alpha,\gamma}$ equals $q_{0}(\gamma)$.
With time independent Markovian feedback protocol, the convergence is exponential with a leading rate $\overline{S}(\Upsilon|\alpha)$.

\subsection{Convergence rate tuning}

Most of the time, when performing a measurement, one prefers it to
take as little time as possible. The use of different partial measurement
methods allows us to tune the convergence rate. Let us take an example.
Suppose we want to discriminate between three possible pointer states
of a system, and suppose that the partial measurement
methods give only ${\tt True}/{\tt False}$ as possible outputs. We
denote $T,F$ the partial measurement results and $1,2,3$ the pointer
states. Each partial measurement method can be tuned to maximize,
up to a measurement error $\varepsilon\ll1$, the probability of one
of its outcome knowing the system is in one of the three pointer states.
We shall show that this is not enough to maximize all convergence rates for arbitrary
limit pointer state. It is the use of different measurement
methods picked randomly that allows us to overcome this convergence rate
problem.

Let us consider for instance two measurement methods. The first one,
denoted $a$, has conditioned probabilities \begin{align*}
p^{a}(T|1)=\varepsilon,p^{a}(T|2)=q,p^{a}(T|3)=1-\varepsilon\end{align*}
 with $q=O(1)$. The second one, denoted $b$, is obtained by switching
the probability conditioned on $1$ and $2$, that is \begin{align*}
p^{b}(T|1)=q,p^{b}(T|2)=\varepsilon,p^{b}(T|3)=1-\varepsilon\end{align*}
 Let us now look at the convergence rate conditioned on the limit
pointer state to be $1$. These are coded in the relative entropies.
If only the measurement method $a$ is used, one has: 
\begin{align*}
S^{a}(1|2) & =\varepsilon\ln\left[\frac{\varepsilon}{q}\right]+(1-\varepsilon)\ln\left[\frac{1-\varepsilon}{1-q}\right]\sim-\ln[1-q]=O(1)\\
S^{a}(1|3) & =\varepsilon\ln\left[\frac{\varepsilon}{1-\varepsilon}\right]+(1-\varepsilon)\ln\left[\frac{1-\varepsilon}{\varepsilon}\right]\sim-\ln[\epsilon]\gg1
\end{align*}
 The convergence of $Q_{n}(3)$ toward $0$ is quick but the one of
$Q_{n}(2)$ is rather slow. If measurement method $b$ is used the
interesting relative entropies are now 
\begin{align*}
S^{b}(1|2) & =q\ln\left[\frac{q}{\varepsilon}\right]+(1-q)\ln\left[\frac{1-q}{1-\varepsilon}\right]\sim-\ln[\varepsilon]\\
S^{b}(1|3) & =q\ln\left[\frac{q}{1-\varepsilon}\right]+(1-q)\ln\left[\frac{1-q}{\varepsilon}\right]\sim-\ln[\varepsilon]
\end{align*}
 All the convergences rates are then high if the limit pointer state
is $1$. But if the limit pointer state is not $1$ but $2$ then,
using only the measurement method $b$, the relative entropy $S^{b}(2|1)$
is 
\[
S^{b}(2|1)=\varepsilon\ln\left[\frac{\varepsilon}{q}\right]
+(1-\varepsilon)\ln\left[\frac{1-\varepsilon}{1-q}\right]\sim-\ln[q]=O(1)
\]
 and the convergence rate toward $2$ is slow.

Now, if at each time one of the two measurement methods is used with
equal probability $\frac{1}{2}$. The convergence rate for any $i,j$
with $i\ne j$ is 
\begin{align*}
\overline{S}(i|j) & =\frac{1}{2}(S^{a}(i|j)+S^{b}(i|j))\sim-\ln[\varepsilon]\gg1
\end{align*}
 As a consequence, the convergence rate is always high, whichever
the limit pointer state is.

In the toy model, if the first partial measurement method
correspond to $\theta-\theta'=\frac{\pi}{3}$, then 
\[
S^{\frac{\pi}{3}}(0|3)\sim0.116
\]
 This is the slowest of all convergence rates. If the partial measurement
method with $\theta-\theta'=\frac{\pi}{6}$ is introduced and the
partial measurement methods are chosen with equal probabilities each
time, then 
\[
\overline{S}(0|3)\sim1.18
\]
 and the slowest of all convergence rates is \[
\overline{S}(1|3)\sim1.10\]
 If only the $\frac{\pi}{3}$ measure is used and the limit pointer
state is $0$, then a theoretical $99\%$ confidence level is reached
after about 50 measures. With the use of the two different partial
measurement methods the same confidence level for the same limit state
is reached in 5 measures. The same number of measurements is needed
if the limit pointer state is $1,2$ or $3$.

\section{Degeneracy and limit quantum state}\label{sec:degeneracy}

Often the quantity we measure is a property common to several pointer
states. In the quantum case, this corresponds to a degenerate projective
Von Neumann measurement. There, at least two different eigenstates share the same eigenvalue.
For our measurement process, degeneracies happen when several distributions $p^{o}(\cdot|\cdot)$ are equal
for different pointer states, so that some states cannot be distinguished. For example, in our toy model,
whatever $\theta-\theta'$ is, we have $p^{\theta-\theta'}(\pm|p)=p^{\theta-\theta'}(\pm|p+4k)$
with $k$ an integer. The pointer state with $p$ photons cannot be distinguished from the one with $p+4k$ photons.

In this section we study the system state evolution when degenerate
repeated partial measurements are performed. In a first part we show
the convergence of the system pointer state distribution.
In a second part we focus on the quantum case and the influence of
phases introduced between degenerate states by the repeated partial
measurement process.

We shall partition the set of configurations into sectors. Let us
define an equivalence relation among pointers by identifying
two pointers whose partial measurement distributions are identical.
That is: two pointers $\alpha$ and $\beta$ are said to be
equivalent (denoted $\alpha \sim \beta$) if, for any partial measurement method $o$ and result
$i$, 
\[
p^{o}(i|\alpha)=p^{o}(i|\beta).
\]
 By definition the sector $\underline{\alpha}$ is the equivalence
class of $\alpha$. In the toy model the sectors are the sets $\underline{p}=\{p+4k,k\in\mathbb{N}\}$
with $p=0,1,2,3$.

\subsection{State distribution convergence}

\label{sec:state_distrib_conv}

We first look at the convergence of the pointer state distribution
$Q_{n}(\cdot)$ in case of degeneracy. The system distributions $Q_{n}(\cdot)$
induce probability distributions $\bar{Q}_{n}(\cdot)$ on sectors
by 
\[
\bar{Q}_{n}(\underline{\alpha}):=\sum_{\alpha'\in\underline{\alpha}}Q_{n}(\alpha').
\]
 Since sectors form a partition of the set of pointer states, we have
$\sum_{\underline{\alpha}}Q_{n}(\underline{\alpha})=1$. The initial
probability of a sector is $\bar{q}_{0}(\underline{\alpha})=\sum_{\alpha'\in\underline{\alpha}}q_{0}(\alpha')$.
The recursion relation (\ref{recurqn}) can obviously be lifted to
a recursion relation for the sector distributions, 
\[
\bar{Q}_{n+1}(\underline{\alpha})=\bar{Q}_{n}(\underline{\alpha})\,\frac{p^{o_{n}}(i_{n}|\alpha)}{\sum_{\underline{\beta}}\bar{Q}_{n}(\underline{\beta})p^{o_{n}}(i_{n}|\beta)}.
\]
 It is identical in structure to eq.(\ref{recurqn}) but with the
bonus that it now is non degenerate. Lets assume that for two different sectors, it exists at least one $\mathbb P$-recurrent partial measurement method distinguishing between the two sectors: if $\alpha \not\sim \beta$, it exists $o\in \mathcal O_s$ and $i \in {\rm spec}(o)$ such that $p^o(i|\alpha)\ne p^o(i|\beta)$.
Thus we can use the non-degenerate case results but applied to the sector distribution.
Hence, $\bar{Q}_{n}(\cdot)$ almost surely converge and \[
\bar{Q}(\underline{\alpha})=\delta_{\underline{\alpha},\underline{\Upsilon}}\]
 with $\underline{\Upsilon}$ the realization dependent limit sector.
The probability that the limit sector be $\underline{\gamma}$ is
equal to $\bar{q}_{0}(\underline{\gamma})$.

  From the martingale property, the state distribution converge (not
only the sector distribution), and the point which remains to be
discussed is what is this limit. Thanks to the relation $q_{0}(\alpha')\, Q_{n}(\alpha)=q_{0}(\alpha)\, Q_{n}(\alpha')$
valid for any $n$ if $\alpha\sim\alpha'$, we shall show that this
limit is \begin{eqnarray}
Q(\alpha)=\left\{ \begin{array}{ll}
0 & \mbox{if }\alpha\not\in\underline{\Upsilon}\\
{q_{0}(\alpha)}/{\bar{q}_{0}(\underline{\Upsilon})} & \mbox{if }\alpha\in\underline{\Upsilon}\end{array}\right.\label{qdegener}\end{eqnarray}
 with $\underline{\Upsilon}$ the limit sector.

Indeed, the state distribution satisfies the recursion relation (\ref{recurqn}).
Thus, if $(i_{k})_{k=0,\cdots,n-1}$ are the $n$ first
partial measurements results, one has \[
\frac{Q_{n}(\alpha')}{q_{0}(\alpha')}=\frac{\prod_{k=1}^{n}p^{o_{k}}(i_{k}|\alpha)}{\sum_{\beta}q_{0}(\beta)\prod_{k=1}^{n}p^{o_{k}}(i_{k}|\beta)}\]
 for any $\alpha'$ in the sector $\underline{\alpha}$. The right
hand side only depends on the sector $\underline{\alpha}$, and thus
$\frac{Q_{n}(\alpha)}{q_{0}(\alpha)}=\frac{Q_{n}(\alpha')}{q_{0}(\alpha')}$
if $\alpha'\in\underline{\alpha}$. From this equality, it follows
that \[
Q(\alpha')=Q(\alpha)\frac{q_{0}(\alpha')}{q_{0}(\alpha)}\]
Since, $\bar{Q}(\underline{\alpha})=\delta_{\underline{\alpha},\underline{\Upsilon}}$,
we have $Q(\alpha)=0$ if $\alpha\notin\underline{\Upsilon}$ and
$1=\sum_{\alpha'\in\underline{\Upsilon}}Q(\alpha')=\frac{Q(\alpha)}{q_{0}(\alpha)}\bar{q}_{0}(\underline{\Upsilon})$,
for any $\alpha\in\underline{\Upsilon}$. Hence, \[
Q(\alpha)=\frac{q_{0}(\alpha)}{\bar{q}_{0}(\underline{\Upsilon})},\quad{\rm for}\ \alpha\in\underline{\Upsilon}\]

The probability of convergence to a sector as well as the limit state distribution (\ref{qdegener}) coincide, in quantum mechanics, with what would have been predicted by Von Neumann rules for degenerate
projective measurements. The approach we have been following so far,
based on tools from classical probability theory, gives no information
on the convergence of the density matrix off-diagonal elements. It is the next section's purpose to discuss the evolution of the density matrix $\rho_{n}$ and not only the evolution of the
probabilities $Q_{n}(\alpha)=\langle\alpha\vert\rho_{n}\vert\alpha\rangle$.

\subsection{Density matrix convergence}

\label{ssec:densitymatconv}

We are now interested in the convergence of the system density matrix. In the basis
of pointer states we may write: 
\[
\rho_{n}=\sum_{\alpha,\beta}A_{n}(\alpha,\beta)\vert\alpha\rangle\langle\beta\vert
\]
 with $A_{n}(\alpha,\alpha)=Q_{n}(\alpha)$. It evolves according
to the recursion relation (\ref{recurrho}). The processes $A_{n}(\alpha,\beta)$,
are not martingales, and their convergence can not be obtain through the
martingale convergence theorem. Actually, they do not always converge. 
To obtain convergence a unitary evolution process has to be subtracted.

For each POVM, a phase between the pointer states is introduced by
the operators $M_{i}^{(o)}$. Even inside a sector this phase can
be nonzero. This possibility comes from the degeneracy criteria we
unraveled previously. Two pointer states, $\alpha,\beta$ can have a nonzero limit probability if they are in the same sector : $\alpha \sim \beta$. This criterion implies a norm equality $|M^{(o)}(i|\alpha)|=|M^{(o)}(i|\beta)|$
for any $i$ in the spectrum of any partial measurement, but not a full equality. So $M^{(o)}(i|\alpha)$ and $M^{(o)}(i|\beta)$ can differ by a phase. The density matrix converges either if this
phase can be set to zero or if we absorb it through a transformation of the evolution.

Let us write the operators $M^{(o)}(i|\alpha)$ in a phase times norm form 
\[
M^{(o)}(i|\alpha)=e^{-i\Delta t(E_{\alpha}+\theta^{(o)}(i|\alpha))}\sqrt{p^{o}(i|\alpha)}
\]
 The specific form of the phase is inspired by the Hamiltonian 
 $H^{(o)}=\sum_{\alpha}(E_{\alpha}\mathbb{I}+H_{p}^{(o)}+H_{\alpha}^{(o)})\vert\alpha\rangle\langle\alpha\vert$.
This is the most general Hamiltonian if one
want $U$ to fulfill the non demolition condition (\ref{Upointer}).
In the above formula, $\Delta t$ is the interaction time between the probe and the system.

Let us define a unitary operator process, diagonal in the pointer state basis,
\[
\widetilde{U}_{n}=\sum_{\alpha}e^{-i\Delta t(nE_{\alpha}
+\sum_{(o,i)\in E}\theta^{(o)}(i|\alpha)N_{n}(o,i))}\vert\alpha\rangle\langle\alpha\vert
\]
 and the unitary equivalent conjugate density matrix process 
 \begin{eqnarray}
\widetilde{\rho}_{n}=\widetilde{U}_{n}^{\dagger}\rho_{n}\widetilde{U}_{n}\label{eq:rho_unitarily_transformed}
\end{eqnarray}
 The diagonal elements of $\rho_{n}$ in the basis $\{\vert\alpha\rangle\}$,
are not affected by this transformation. Their limits stay the same.
Thus if $\alpha$ or $\beta$ are not in the limit sector $\underline{\Upsilon}$,
according to the Cauchy-Schwartz theorem, $\widetilde{A}_{\infty}(\alpha,\beta)=0$.
We are then interested in the limit of the elements $\widetilde{A}_{n}(\alpha,\beta)$
with $\alpha,\beta\in\underline{\Upsilon}$.

If $\beta\in\underline{\alpha}$, then 
$q_{0}(\alpha)\widetilde{A}_{n}(\alpha,\beta)=a_{0}(\alpha,\beta)Q_{n}(\alpha)$.
Repeating the discussion made in the section \ref{sec:state_distrib_conv}, we get
\begin{eqnarray*}
\widetilde{A}_{\infty}(\alpha,\beta)=\left\{ \begin{array}{ll}
a_{0}(\alpha,\beta)/\overline{q}_{0}(\underline{\Upsilon}) & \mbox{if }\alpha,\beta\in\underline{\Upsilon}\\
0 & \mbox{else}\end{array}\right.\end{eqnarray*}

Hence, $\widetilde{\rho}_{n}$ has an almost sure limit which coincides
with the result of a Von Neumann measurement: $\widetilde{\rho}_{\infty}$ is equal to $\rho_0$ projected
on the system subspace corresponding to the sector
$\underline{\Upsilon}$. 
\begin{eqnarray}
\lim_{n\to\infty}\widetilde{\rho}_{n}=\frac{1}{\overline{q}_{0}(\underline{\Upsilon})}\mathrm{P}_{\underline{\Upsilon}}\rho_{0}\mathrm{P}_{\underline{\Upsilon}}\label{eq:limit_discrete_rhotilde}
\end{eqnarray}
where $\mathrm{P}_{\underline{\Upsilon}}:=\sum_{\gamma\in\underline{\Upsilon}}\vert\gamma\rangle\langle\gamma\vert$
is the projector on the subspace corresponding to the sector $\underline{\Upsilon}$.

In some cases the unitary operator process can be reduced to a deterministic
one. For example if $\theta^{(o)}(i|\alpha)=\theta^{(o)}(i|\beta)$
for every $(o,i)\in E$ and every $\alpha\sim\beta$, then we can
chose $\widetilde{U}_{n}=e^{-i\Delta tH_{s}}$. 
This is the case in the toy model, from the periodicity of trigonometric
functions, in a given sector all phases introduced by partial measurements
are equal. If the limit sector is $\underline{p}=\{p+4k,k\in\mathbb{N}\}$,
then the photon field limit state will be 
\[
\lim_{n\to\infty}e^{in\Delta t\hat{p}}\rho_{n}\, e^{-in\Delta t\hat{p}}=\sum_{k,k'\in\mathbb{N}}\frac{a_{0}(p+4k,p+4k')}{\bar{q}_{0}(\underline{p})}\,\vert p+4k\rangle\langle p+4k'\vert
\]
 One other example corresponds to the case where the phases $\theta^{(o)}(i|\alpha)$
do not depend on $(o,i)$ but depend on $\alpha$. Then one can define
$H_{eff.}=H_{s}+\sum_{\alpha}\theta(\alpha)\vert\alpha\rangle\langle\alpha\vert$
and $\widetilde{U}_{n}=e^{-in\Delta tH_{eff.}}$. 

In most of the cases the unitary evolution $\widetilde{U}_{n}$ is
a stochastic process and then in the limit $n\to\infty$, it remains
a stochastic rotation inside the limit sector.
When $\widetilde{U}_n$ is deterministic, the remaining rotation is deterministic too.

\section{Continuous diffusive limit}\label{sec:continuous time}

We shall now prove the convergence of the discrete processes we consider toward
 processes driven by time continuous Belavkin diffusive equations. Our proof, 
different from that used in \cite{pellegrini_diff},
allows us to derive the continuous measurement diffusive equation
not only for the quantum repeated indirect measurement process but
also for the macroscopic Bayesian apparatus we defined. The quantum
case is a peculiar realization of it. 

The time continuous equation is found as a scaling limit of the discrete evolution
when $n$ goes to infinity with $t=n\delta$ fixed ($\delta=\Delta t$). We first study
the pointer state distribution scaling limit, $Q_{t}(\alpha):=\lim_{\delta\to0}Q_{[t/\delta]}(\alpha)$.
The evolution equation for $Q_{t}(\alpha)$ is given in
eq.(\ref{stoQ}) below. We then look at the time continuous limit
of the density matrix evolution and get the Belavkin diffusive
equation eq.(\ref{def:belavkin_eq}). In the quantum case, the continuous
limit requires rescaling appropriately the system-probe interaction
Hamiltonian as \begin{eqnarray}
H=H_{s}\otimes\mathbb{I}_{p}+\mathbb{I}_{s}\otimes H_{p}+\frac{1}{\sqrt{\delta}}H_{I}\label{Hinter}\end{eqnarray}
 We present in some details the case with a unique partial measurement
method. The results are then easily extended to cases with different
measurement methods.

\subsection{Continuous time limit of the pointer state distribution}

We are here interested in the state distribution continuous time limit.
The results presented in this section apply to the general Bayesian
recursion relation (\ref{recurqn}) --- which in particular includes
the case of repeated QND measurements. To begin with, we assume that
there is only one partial measurement method%
\footnote{This condition will be relaxed in section \ref{sssec:diff-part-meas}.%
}. Henceforth we suppress $o$ from all the notations and let $\mathcal I$
stand for the index set of outcomes. Note that the two filtrations
${\mathcal{F}}_{n}$ and ${\mathcal{F}}'_{n}$ coincide and carry
the information on the first $n$ partial measurements.

We assume that the conditional probabilities $p(i|\alpha)$ depend
on a further small parameter $\delta$, and are of the form 
\begin{equation}
p(i|\alpha)=p_{0}(i)(1+\sqrt{\delta}\,\Gamma_{\delta}(i|\alpha))\label{eq:pdelta}
\end{equation}
 with $p_{0}(i)>0$ for all $i$'s and that $\Gamma(i|\alpha):=\lim_{\delta\rightarrow0^{+}}\Gamma_{\delta}(i|\alpha)$
exists. Then $\sum_{i}p_{0}(i)=1$, so that the $p_{0}(i)$'s specify
a probability measure, and for every $\delta$, $\sum_{i}p_{0}(i)\Gamma_{\delta}(i|\alpha)=0$.
The important point is that $p_{0}(i)$ is independent of $\alpha$.

These hypothesis are of course satisfied in the quantum case with
QND interaction Hamiltonian $H_{I}=\sum_{\alpha}\vert\alpha\rangle\langle\alpha\vert\otimes H_{\alpha}$ and rescaling $H_I \to \frac{1}{\sqrt{\delta}}H_I$. Then
$p_{0}(i)=|\langle i|\Psi\rangle|^{2}$ is the probability measure
in absence of interaction, and \[
\Gamma(i|\alpha):=2\,{\rm Im}\Big(\frac{\langle i\vert H_{\alpha}\vert\Psi\rangle}{\langle i\vert\Psi\rangle}\Big)\]
 We assume that ${\langle i\vert\Psi\rangle}\not=0$ for all $i$.

We first need to make precise the sense in which
a limit on $Q_{n}(\alpha)$ is to be taken.

For a fixed $\omega\in\Omega$ the limit
$\lim_{\delta\to0}Q_{[t/\delta]}(\alpha)$ is not expected to
exist\footnote{Think of the simple random walk: the convergence to Brownian
  motion is not sample by sample because $\frac{S_{2n}}{\sqrt{2n}}$ has no
  reason to be close to $\frac{S_{n}}{\sqrt{n}}$.}. But there is some hope that,
properly defined, a limit for the law of the process $Q_{[t/\delta]}(\alpha)$,
$t\in{\mathbb{R}}^{+}$ exists. We shall prove in Appendix \ref{app:conv} that
this is the case, specifying a bit the kind of convergence that is involved.
Under the limiting law, the process $Q_{t}(\cdot)$ satisfies the stochastic
equation 
\begin{eqnarray}
  dQ_{t}(\alpha)=Q_{t}(\alpha)\sum_{i}\big(\Gamma(i|\alpha)-\langle\Gamma_{i}\rangle_{t}\big)\,
  dX_{t}(i)
  \label{stoQ}\end{eqnarray} 
  where \[
\langle\Gamma_{i}\rangle_{t}:=\sum_{\beta}Q_{t}(\beta)\Gamma(i|\beta).\] Here
$X_{t}(i)$, with $\sum_{i}X_{t}(i)=0$, are continuous martingales with quadratic
covariation \begin{eqnarray} dX_{t}(i)dX_{t}(j)=dt\,\big(\delta_{i,j}\,
  p_{0}(i)-p_{0}(i)p_{0}(j)\big).\label{quadraX}\end{eqnarray}

We shall show that a vector solving this equation is a bounded martingale,
to which the martingale convergence theorem can be applied with results
similar to those in the discrete case:
\[
Q_{\infty}(\alpha)=\left\{ \begin{array}{ll}
0 & \mbox{ if }\alpha\not\in\underline{\Upsilon}\\
{q_{0}(\alpha)}/{\bar{q}_{0}(\underline{\alpha})} & \mbox{ if }\alpha\in\underline{\Upsilon}\end{array}\right.\]
 with $\underline{\Upsilon}$ the limit sector. However, the sector definition is not the same as in the discrete case. In time continuous,
$\alpha$ and $\beta$ are in the same sector if and only if $\Gamma(i|\alpha)=\Gamma(i|\beta)$
for all partial measurement result $i$. The probability for the system
to be in the sector $\underline{\alpha}$ in the limit $t$ goes to
infinity is $\bar{q}_{0}(\underline{\alpha})=\sum_{\alpha'\in \underline{\alpha}} q_0(\alpha')$.

The convergence is still exponential \[
Q_{t}(\alpha)=\exp{(-{t}/{\tau_{\Upsilon\alpha}})},\quad\mbox{ if }\alpha\not\in\underline{\Upsilon}\]
 with characteristic convergence time $\tau_{\gamma\alpha}$, \begin{eqnarray}
2/\tau_{\gamma\alpha}={\sum_{i}p_{0}(i)\big(\Gamma(i|\alpha)-\Gamma(i|\gamma)\big)^{2}}\label{deftau}\end{eqnarray}
 This coincides with the convergence rate we would have found by taking
the relative entropy $S(\gamma|\alpha)$ scaling limit. However,
it is somewhat difficult to decipher that it originates from a relative
entropy by only knowing its expression in the continuous-time limit.

\subsubsection{Preparation}

\label{sec:prep}

We work with the model $(\Omega,{\mathcal{F}},{\mathbb{P}})$.

Our derivation is based on the use of the counting processes $N_{n}(i)$.
Recall that $N_{0}(i)=0$ and that $N_{n}(i):=\sum_{1\leq m\leq n}\epsilon_{m}(i)$
for $n\geq1$, where $\epsilon_{n}(i):={\mathbb{1}}_{I_{n}=i}$ is
$1$ if the ${\rm n}^{\rm th}$ partial measurement outcome is $i$ and $0$ otherwise.

We start by listing some properties of these counting processes and
their relationship to the solution of (\ref{recurqn}). Then we shall
formulate and prove the analogous statements for the continuous time
limit.

It is obvious that the filtration ${\mathcal{F}}_{0},{\mathcal{F}}_{1},\cdots$
is the natural filtration of the vector counting processes $N_{n}$.

Also recall that the random recursion relation (\ref{recurqn}) can
be solved in terms of the counting processes as 
\[
Q_{n}(\alpha)=q_{0}(\alpha)\frac{\prod_{i}p(i|\alpha)^{N_{n}(i)}}{\sum_{\beta}q_{0}(\beta)\prod_{i}p(i|\beta)^{N_{n}(i)}}.
\]

A trivial but crucial observation is that under each ${\mathbb{P}}_{\alpha}$,
$N_{n}$ is the sum of independent identically distributed (i.i.d) random vectors.

As a first consequence, a simple computation leads to \begin{eqnarray}
{\mathbb{E}}\left(e^{\sum_{l=1}^{k}\sum_{i}\lambda_{l}(i)(N_{n_{l}}(i)-N_{n_{l-1}}(i))}\right) & =\nonumber \\
 &  & \hspace{-2cm}\sum_{\alpha}q_{0}(\alpha)\prod_{l=1}^{k}\left(\sum_{i}e^{\lambda_{l}(i)}p(i|\alpha)\right)^{n_{l}-n_{l-1}}\label{eq:charfunc}\end{eqnarray}
 for $k\geq1$, arbitrary non-decreasing sequences of integers $0=n_{0}\leq n_{1}\leq\cdots\leq n_{k}$
of length $k$, and arbitrary (complex) $\lambda_{l}(i)$'s. A second
consequence is that under ${\mathbb{P}}_{\alpha}$ each $N_{n}(i)$
is a sub-martingale and $N_{n}(i)=(N_{n}(i)-np(i|\alpha))+np(i|\alpha)$
is its Doob decomposition as a martingale plus a predictable (in that
case deterministic) increasing process. Moreover, if $n\geq1$, and
if $X$ is an ${\mathcal{F}}_{n-1}$ measurable random variable, we
compute 
\begin{eqnarray*}
{\mathbb{E}}(X\epsilon_{n}(i)) & = & \sum_{\alpha}q_{0}(\alpha){\mathbb{E}}_{\alpha}(X\epsilon_{n}(i))\\
 & = & \sum_{\alpha}q_{0}(\alpha){\mathbb{E}}_{\alpha}(X)p(i|\alpha)\;\;=\;\;\sum_{\alpha}{\mathbb{E}}(XQ_{n-1}(\alpha))p(i|\alpha).
\end{eqnarray*}
 For the last equality we used the $Q$'s characterization 
as Radon-Nikodym derivatives. This proves that ${\mathbb{E}}(\epsilon_{n}(i)|{\mathcal{F}}_{n-1})=\sum_{\alpha}Q_{n-1}(\alpha)p(i|\alpha)=\pi_{n}(i)$.
Hence, setting 
\[A_{n}(i):=\sum_{m=1}^{n}\pi_{m}(i),\] an increasing predictable process, we
find that $X_{n}(i):=N_{n}(i)-A_{n}(i)$ is an ${\mathcal{F}}_{n}$-martingale
under ${\mathbb{P}}$, so each $N_{n}(i)$ is again a sub-martingale with Doob
decomposition \begin{equation} N_{n}(i)=X_{n}(i)+A_{n}(i)\end{equation} under
${\mathbb{P}}$.

Finally, by some simple algebra we may rephrase the random recursion relation satisfied by the $Q$'s as a stochastic difference equation
\begin{equation}
Q_{n}(\alpha)-Q_{n-1}(\alpha)=Q_{n-1}(\alpha)\sum_{i}\frac{p(i|\alpha)}{\pi_{n-1}(i)}(X_{n}(i)-X_{n-1}(i)).\label{stoq}\end{equation}

\subsubsection{Derivation of the pointer state distribution evolution}
\label{sssec:deriv_continuous_distrib_evol}

Equation (\ref{stoq}) admits eq.(\ref{stoQ}) as a naive continuous time limit when
$\delta$, the scaling parameter, goes to $0^+$. To put
the validity of this formal approach on a firmer ground, one needs to prove the
existence of a continuous time limit. This is a classical topic, but the
presence of the scaling parameter $\delta$ in various places prevents us from
applying standard theorems straightforwardly. So we rely on a down-to-earth
approach, which is rather technical. For this reason we relegated the argument
to appendix \ref{app:conv}. This is where the interested reader should look for
some background, precise definitions, etc. We give here a brief summary:

-- By an appropriate interpolation procedure, one defines a $\delta$-dependent
push-forward $\mu_{\alpha}(\delta)$ of each ${\mathbb P}_\alpha$ and
$\mu(\delta)$ of ${\mathbb P}$ in $C_{0}({\mathbb{R}}^{+},{\mathbb{R}}^{I})$, the
space of continuous functions from ${\mathbb{R}}^{+}$ to ${\mathbb{R}}^{I}$
vanishing at $0$.

-- We are not able to prove the convergence in law of the $\mu_{\alpha}(\delta)$
or of $\mu(\delta)$ when $\delta \rightarrow 0^+$.

-- However, the finite dimensional distributions of the joint processes $N_n(i)$
and $Q_n(\alpha)$ under each ${\mathbb P}_{\bullet}$ (where $\bullet$ stands
either for an element of ${\mathcal S}$ or for nothing) admit, after appropriate
time dependent centering and scaling, continuous time limits which are the joint
finite dimensional distributions, under a probability measure $\mu_{\bullet}$ on
$C_{0}({\mathbb{R}}^{+},{\mathbb{R}}^{I})$, for processes $W_t(i)$, to be
thought of as
\[
\lim_{\delta \rightarrow 0^+}  \sqrt{\delta}(N_{t/\delta}(i)-p_{0}(i)t/\delta),
\]
and $Q_t(\alpha)$, to be thought of as $\lim_{\delta \rightarrow 0^+}
Q_{t/\delta}(\alpha)$. 

-- The process $W_t$ is the canonical coordinate process on
$C({\mathbb{R}}^{+},{\mathbb{R}}^{I})$, and its natural filtration
$\mathcal{G}_{t}$ is to be thought of as the continuous time limit of the
natural filtration for $N_n$, i.e. as the information collected by indirect
measurements up to time $t$.

-- The identity $\mu=\sum_{\alpha} q_0(\alpha) \mu_{\alpha}$ holds. The
Radon-Nikodym derivative of $\mu(\alpha)$ with respect to $\mu$ on
$\mathcal{G}_{t}$ is ${M_{t}(\alpha)}/{M_{t}}$ where
\[
M_{t}(\alpha):=e^{\sum_{i}\Gamma(i|\alpha)W_{t}(i)-\frac{t}{2}
  \sum_{i}p_{0}(i)\Gamma(i|\alpha)^{2}},
\qquad M_{t}:=\sum_{\alpha}q_{0}(\alpha)M_{t}(\alpha)\]
For each $\alpha$, $M_{t}^{-1}(\alpha)$ is a $\mathcal{G}_{t}$-martingale under
$\mu_{\alpha}$, and $M_{t}^{-1}$ is a $\mathcal{G}_{t}$-martingale under
$\mu$. 

-- For each $T>0$, under the measure $M_T^{-1}d\mu$ (which coincides with
$M_T^{-1}(\alpha)d\mu_{\alpha}$ for every $\alpha$), the process $(W_t)_{t\in
  [0,T]}$ is a continuous time-homogeneous centered Gaussian process with
  covariance $\min(t,s) \big(\delta_{i,j}\, p_{0}(i)-p_{0}(i)p_{0}(j)\big)$.
  Thus, by Girsanov's theorem, under each $\mu_{\alpha}$, $W_t$ is a continuous
  time-homogeneous Gaussian process with independent increments, characterized
  by
\begin{eqnarray*} {\mathbb{E}}^{\mu_{\alpha}}(W_t(i)) & = &
  tp_{0}(i)\Gamma(i|\alpha) \\ 
\text{Cov}^{\mu_{\alpha}}(W_t(i),W_s(j)) & = & \min(t,s) \big(\delta_{i,j}\,
p_{0}(i)-p_{0}(i)p_{0}(j)\big). \end{eqnarray*} 

--  There is an explicit formula for the $Q_{t}$'s in terms of the $W_{t}$'s,
namely: 
\[
Q_{t}(\alpha)= q_{0}(\alpha) \frac{M_{t}(\alpha)}{M_{t}} =
  q_{0}(\alpha)\frac{e^{\sum_{i}\Gamma(i|\alpha)
      W_{t}(i)-\frac{t}{2}\sum_{i}p_{0}(i)\Gamma(i|\alpha)^{2}}}
  {\sum_{\beta}q_{0}(\beta)e^{\sum_{i}\Gamma(i|\beta)
      W_{t}(i)-\frac{t}{2}\sum_{i}p_{0}(i)\Gamma(i|\beta)^{2}}}.
\] 

We are now in position to check that all the properties established
in the discrete setting, as listed in section \ref{sec:prep}, have
a direct naive counterpart in the continuous time setting.

The construction of the filtration $\mathcal{G}_{t}$ as the natural filtration
for the canonical process was already explained. We have also already mentioned
that there is an explicit formula for the $Q_{t}$'s. The counterpart of
(\ref{eq:charfunc}), the Laplace transform of the
counting processes joint distributions is given for the canonical process in eq.(\ref{eq:charfunc5}), Appendix \ref{app:conv}.

The counterpart of the counting process Doob decomposition under
${\mathbb{P}}_{\alpha}$ is
$W_{t}(i)=(W_{t}(i)-tp_{0}(i)\Gamma(i|\alpha))+tp_{0}(i)\Gamma(i|\alpha)$ under
$\mu_{\alpha}$.

To get the counterpart of the counting process Doob-Meyer decomposition under
${\mathbb{P}}$, i.e. the Doob-Meyer decomposition of $W_{t}(i)$ under $\mu$, we
use Girsanov's theorem. As recalled above, for every $T >0$
$(W_{t}(i))_{t\in [0,T]}$ is a continuous martingale under $M_T^{-1}d\mu$. From 
\[
dM_{t}/M_{t}=\sum_{\alpha}q_{0}(\alpha)\frac{M_{t}(\alpha)}{M_{t}}\sum_i\Gamma(i|\alpha)dW_{t}(i)=\sum_{\alpha}Q_{t}(\alpha)\sum_i\Gamma(i|\alpha)dW_{t}(i),
\]
we infer that the increasing process
\[A_{t}(i):=\int_{0}^{t}ds\,\sum_{\alpha}Q_{s}(\alpha)p_{0}(i)\Gamma(i|\alpha)\]
is the compensator of $W_{t}(i)$, i.e. 
\[X_{t}(i):=W_{t}(i)-A_{t}(i)\]
is a ${\mathcal{G}}_{t}$ martingale under $\mu$, with quadratic
variation given by (\ref{quadraX}). It is easily seen that $A_{t}(i),X_{t}(i)$
are the obvious continuous time limits of $A_{n}(i),X_{n}(i)$.

It remains to write down the stochastic evolution equations for the
$Q_{t}$'s. By It\^o's formula for a ratio, we find \[
\frac{dQ_{t}(\alpha)}{Q_{t}(\alpha)}=\left(\frac{dM_{t}(\alpha)}{M_{t}(\alpha)}-\frac{dM_{t}}{M_{t}}\right)\left(1-\frac{dM_{t}}{M_{t}}\right).\]
 leading immediately to (\ref{stoQ}) which we reproduce for convenience:
\[
dQ_{t}(\alpha)=Q_{t}(\alpha)\sum_{i}\big(\Gamma(i|\alpha)-\langle\Gamma_{i}\rangle_{t}\big)\, dX_{t}(i)\]
where $\langle\Gamma_{i}\rangle_{t}:=\sum_{\beta}Q_{t}(\beta)\Gamma(i|\beta)$.
Note again that this equation is also the naive continuous time limit of
the discrete equation (\ref{stoq}).

\vspace{0.3cm}

To summarize, one makes no mistakes if one works naively and forgets
about the lengthy rigorous construction of the continuous time
limit. This gives us confidence in what follows to proceed straightforwardly
in the derivation of continuous time equations in more complicated
situations.

\subsubsection{Convergence of the continuous time evolution}
\label{sssec:conv_conttime_distrib}

We now prove the convergence of $Q_{t}(\alpha)$ when $t$ goes to
infinity. Its almost sure convergence is a direct consequence of its
martingale property. We need to prove that the final distribution
is \begin{eqnarray}
Q_{\infty}(\alpha)=\left\{ \begin{array}{ll}
0 & \mbox{ if }\alpha\not\in\underline{\Upsilon}\\
{q_{0}(\alpha)}/{\bar{q}_{0}(\underline{\alpha})} & \mbox{ if }\alpha\in\underline{\Upsilon}\end{array}\right.\label{Qlim}\end{eqnarray}
 and that the convergence is exponential with the characteristic time
$\tau_{\Upsilon\alpha}$.

First we prove that the limit of the sector distribution $\bar{Q}_{t}(\underline{\alpha}):=\sum_{\alpha\in\underline{\alpha}}Q_{t}(\alpha)$
is \[
\bar{Q}_{\infty}(\underline{\alpha})=\delta_{\underline{\alpha},\underline{\Upsilon}}\]
 The equation of evolution for the sector distribution is \[
d\bar{Q}_{t}(\underline{\alpha})=\sum_{\alpha'\in\underline{\alpha}}dQ_{t}(\alpha')=\bar{Q}_{t}(\underline{\alpha})\sum_{i}(\Gamma(i|\alpha)-\langle\Gamma_{i}\rangle_{t})dX_{t}(i)\]
 In the limit $t\to\infty$, we have $Q_{\infty}(\underline{\alpha})(\Gamma(i|\alpha)-\langle\Gamma_{i}\rangle_{\infty})=0$
for all $i$, $\mu$-almost surely. Then either $\bar{Q}_{\infty}(\underline{\alpha})=0$
or $\Gamma(i|\alpha)=\sum_{i}\bar{Q}_{\infty}(\underline{\beta})\Gamma(i|\beta)$.
Since $\Gamma(i|\alpha)\ne\Gamma(i|\beta)$ if $\underline{\alpha}\ne\underline{\beta}$,
the solution to the limit equation is $\bar{Q}_{\infty}(\underline{\alpha})=\delta_{\underline{\alpha},\underline{\Upsilon}}.$

Second we show that $\frac{Q_{t}(\alpha')}{Q_{t}(\alpha)}=\frac{q_{0}(\alpha')}{q_{0}(\alpha)}$
if $\alpha',\alpha$ are in the same sector. As in the discrete case,
this relation implies eq.(\ref{Qlim}). We compute: \[
\frac{Q_{t}(\alpha)}{Q_{t}(\alpha')}\, d\frac{Q_{t}(\alpha')}{Q_{t}(\alpha)}=\frac{dQ_{t}(\alpha')}{Q_{t}(\alpha')}-\frac{dQ_{t}(\alpha)}{Q_{t}(\alpha)}+(\frac{dQ_{t}(\alpha)}{Q_{t}(\alpha)})^{2}-\frac{dQ_{t}(\alpha')}{Q_{t}(\alpha')}\frac{dQ_{t}(\alpha)}{Q_{t}(\alpha)}\]
 Since, $\frac{dQ_{t}(\alpha')}{Q_{t}(\alpha')}=\frac{dQ_{t}(\alpha)}{Q_{t}(\alpha)}$
for $\alpha$ and $\alpha'$ in the same sector, we obtain $d\frac{Q_{t}(\alpha')}{Q_{t}(\alpha)}=0$
if $\alpha,\,\alpha'\in\underline{\alpha}$. For all time $t$,
\[
\frac{Q_{t}(\alpha')}{Q_{t}(\alpha)}=\frac{q_{0}(\alpha')}{q_{0}(\alpha)}\mbox{ if }\alpha \sim\alpha'
\]
 which achieves the proof for the limit pointer state distribution.

Finally we prove the exponential convergence. The tools we use are
the convergence of the pointer state distribution and the It\^o
calculus. From the distribution convergence, we have $\langle\Gamma_{i}\rangle_{t}\simeq\Gamma(i|\Upsilon)$
for $t$ large enough. The evolution equation (\ref{stoQ}) for $\alpha\not\in\underline{\Upsilon}$
becomes $dQ_{t}(\alpha)\simeq Q_{t}(\alpha)\sum_{i}(\Gamma(i|\alpha)-\Gamma(i|\Upsilon))dX_{t}(i)$.
This equation is a well known stochastic exponential equation. Thus,
at large time $t$, with good approximation 
\[
Q_{t}(\alpha)\simeq {\rm const.}\,e^{-\frac{t}{2}\sum_{i}p_{0}(i)(\Gamma(i|\alpha)-\Gamma(i|\Upsilon))^{2}+\sum_{i}X_{t}(i)(\Gamma(i|\alpha)-\Gamma(i|\Upsilon))}
\]
 for $\alpha\not\in\underline{\Upsilon}$. Keeping only the leading
term in the exponential we obtain the exponential decrease, $Q_{t}(\alpha)\simeq\exp(-{t}/{\tau_{\Upsilon\alpha}})$,
with $\tau_{\Upsilon\alpha}$ given in eq.(\ref{deftau}).

\subsubsection{Different partial measurement methods}
\label{sssec:diff-part-meas}

The previous results can easily be extended to cases where different measurement methods are randomly used. 
Since proofs are similar to those of previous sections, here we only present a general outline of the approach.
We limit ourselves to a time and realization independent partial measurement
method distribution. In this case $d_{n}(o_{1},i_{1},\cdots,i_{n},o_{n+1})=\prod_{k=1}^{n+1}c(o_{k})$ with $\sum_o c(o)=1$.

To stay in the scope of the diffusive limit we assume that for any
$o$, $\langle i\vert\Psi^{(o)}\rangle\ne0$.

Following previous sections, we define linear interpolations $W_{t}^{(\delta)}(o,i)$
of the counting processes which naively read
\[
\sqrt{\delta}(N_{t/\delta}(o,i)-c(o)p_{0}^{o}(i)t/\delta)
\]
See Appendix \ref{app:detailB} for precise definitions.
As shown in this appendix, all finite dimensional distribution functions of $W_{t}^{(\delta)}(o,i)$ under (a push-forward of) $\mathbb{P}_\alpha$ (resp. $\mathbb{P}$) have a finite limit as $\delta\to 0^+$ which coincide with those of continuous random processes, denoted $W_{t}(o,i)$, under appropriate measures denoted $\mu_\alpha$ (resp. $\mu$). Under $\mu_\alpha$, $W_{t}(o,i)$ is a Gaussian process with
\begin{eqnarray*} 
{\mathbb{E}}^{\mu_{\alpha}}(W_{t}(o,i) & = & t\, c(o)\, p^{o}_{0}(i)\, \Gamma^{(o)}(i|\alpha) \\ 
\text{Cov}^{\mu_{\alpha}}(W_t(o,i),W_s(o',j)) & = & \min(t,s)\, 
\big( c(o)p_{0}^{o}(i)\delta_{(o,i),(o',j)}-c(o)p_{0}^{o}(i)c(o')p_{0}^{o'}(j) \big). 
\end{eqnarray*} 
with $p_{0}^{o}(i)=|\langle i\vert\Psi^{(o)}\rangle|^2$ and 
$\Gamma^{(o)}(i|\alpha)=2\mathrm{Im}\left(\frac{\langle i\vert H_{\alpha}^{(o)}\vert\Psi^{(o)}\rangle}{\langle i\vert\Psi^{(o)}\rangle}\right)$.

The measure $\mu$ is the sum $\mu=\sum_{\alpha} q_0(\alpha) \mu_{\alpha}$. The
Radon-Nikodym derivative of $\mu(\alpha)$ with respect to $\mu$ is ${M_{t}(\alpha)}/{M_{t}}$ 
where $M_t=\sum_\alpha q_0(\alpha) M_t(\alpha)$ with
\[
M_{t}(\alpha)=e^{\sum_{(o,i)\in E}\Gamma^{(o)}(i|\alpha)W_{t}(o,i)-\frac{t}{2}\sum_{(o,i)\in E}c(o)p_{0}^{o}(i)\Gamma^{(o)}(i|\alpha)^{2}}\]
 As in the section \ref{sssec:deriv_continuous_distrib_evol} we define
\[
X_{t}(o,i)=W_{t}(o,i)-\int_{0}^{t}\sum_{\alpha}Q_{s}(\alpha)c(o)p_{0}^{o}(i)\Gamma^{(o)}(i|\alpha)ds
\]
 The $X_{t}(o,i)$ are martingales under $\mu$. From this definition
we obtain straightforwardly 
\[
dQ_{t}(\alpha)=Q_{t}(\alpha)\sum_{(o,i)\in E}(\Gamma^{(o)}(i|\alpha)-\langle\Gamma^{(o)}(i)\rangle_{t})dX_{t}(o,i)
\]
with $\langle \Gamma^{(o)}(i) \rangle_t=\sum_\alpha \Gamma^{(o)}(i|\alpha) Q_t(\alpha)$ and
 \[
dX_{t}(o,i)dX_{t}(o',j)=dt\big(c(o)p_{0}^{o}(i)\delta_{(o,i),(o',j)}-c(o)p_{0}^{o}(i)c(o')p_{0}^{o'}(j)\big)
\]

The limit of $Q_{t}(\alpha)$ is the same but the sectors are now
the sets of basis states such that $\Gamma^{(o)}(i|\alpha')=\Gamma^{(o)}(i|\alpha)$
for all partial measurement methods and all partial measurement results.
The convergence toward the limit distribution is exponential \[
Q_{t}(\alpha)\simeq\exp\Big[-\frac{t}{2}\sum_{(o,i)\in E}c(o)p_{0}^{o}(i)(\Gamma^{(o)}(i|\alpha)-\Gamma^{(o)}(i|\Upsilon))^{2}\Big]\]
 The approximation hold if $t$ is large enough. The convergence is
exponential with a characteristic time \[
\frac{2}{\tau_{\Upsilon\alpha}}=\sum_{(o,i)\in E}c(o)p_{0}^{o}(i)(\Gamma^{(o)}(i|\alpha)-\Gamma^{(o)}(i|\Upsilon))^{2}\]
 We find a convergence rate which is a mean convergence rate as in
the discrete case. The same result is found by taking the scaling limit of the discrete case mean
relative entropy.

\subsection{Density matrix evolution}
\label{sec:densequ}

We are now interested in the density matrix evolution.

As in section \ref{ssec:densitymatconv}, the density matrix at time $n$ can be decomposed in the basis of pointer
states:
\[
\rho_{n}=\sum_{\alpha,\beta}A_{n}(\alpha,\beta)\vert\alpha\rangle\langle\beta\vert\]
The same decomposition applies to the time continuous density matrix we will define. 
The recurrence relation (\ref{recurrho}) translates for $A_{n}(\alpha,\beta)$ in 
\[
A_{n}(\alpha,\beta)=\frac{A_{n-1}(\alpha,\beta)M^{(o_{n})}(i_{n}|\alpha){M^{(o_{n})}(i_{n}|\beta)}^{\star}}{\sum_{\gamma}q_{n-1}(\gamma)p^{o_{n}}(i_{n}|\gamma)}
\]
Where $M^{(o)}(i|\alpha)=\langle i\vert U^{(o)}(\alpha)\vert\Psi^{(o)}\rangle$. For $\alpha=\beta$, this reproduces the pointer state distribution recurrence relation (\ref{recurqn}), as expected.

We first limit ourselves to the case where only one partial measurement method is used and we omit the index $o$. The results will then be generalized to different partial measurement methods.
We used a few hypotheses to get the continuous-time limit:

--  The first two assumptions are related to the development in $\sqrt{\delta}$
of the conditional probabilities $p(i|\alpha)$. As stated before, the interaction Hamiltonian must be rescaled
$H_{I}\to\frac{1}{\sqrt{\delta}}H_{I}$ and for any partial measurement
result $i$, $\langle i\vert\Psi\rangle\ne0$. Then \[
p(i|\alpha)=p_{0}(i)(1+\sqrt{\delta}\,\Gamma_{\delta}(i|\alpha))\]
with $p_{0}(i)=|\langle i|\psi\rangle|^2$. The assumption $\langle i\vert\Psi\rangle\ne0$ leads to the diffusive
limit. If this condition is not fulfilled for every $i$, then a jump-diffusion
limit is found as shown in \cite{pellegrini_benoist_toappear}.

 -- A third assumption is needed to obtain a convergence
of the evolution of the phases between different pointer states. The
interaction Hamiltonian expectation must be zero : 
\[
\langle\Psi\vert H_{I}\vert\Psi\rangle=0
\]

Under these assumptions, we show in Appendix \ref{app:density} that the time continuous evolution derived from
the discrete time case is \begin{align}
A_{t}(\alpha,\beta)=A_{0}(\alpha,\beta)\frac{e^{l(\alpha,\beta)t-i\sum_{i}(c(i|\alpha)-c(i|\beta)^{\star})W_{t}(i)}}{\sum_{\gamma}q_{0}(\gamma)e^{\sum_{i}-i\Gamma(i|\gamma)W_{t}(i)-\frac{t}{2}p_{0}(i)\Gamma(i|\gamma)^{2}}}\label{def:A_t}\end{align}
 with $c(i|\alpha)=\frac{\langle i\vert H_{\alpha}\vert\Psi\rangle}{\langle i\vert\Psi\rangle}$
and \[
l(\alpha,\beta):=-i(E_{\alpha}-E_{\beta})-\frac{1}{2}\sum_{i}p_{0}(i)(|c(i|\alpha)|^{2}+|c(i|\beta)|^{2}-c(i|\alpha)^{2}-{c(i|\beta)^{\star}}^{2})\]
 If we set $\alpha=\beta$ we recover the result on the pointer state
distribution.

A simple computation using It\^o rules shows that this process is solution of a Belavkin diffusive equation:
 \begin{eqnarray}
d\rho_{t}=L(\rho_{t})-i\sum_{i}(C_{i}\rho_{t}-\rho_{t}C_{i}^{\dagger}-\rho_{t}Tr[(C_{i}-C_{i}^{\dagger})\rho_{t}])dX_{t}(i)\label{def:belavkin_eq}
\end{eqnarray}
 with the Lindbladian 
 \[
L(\rho)=-i[H_{s},\rho]+\sum_{i}p_{0}(i)(C_{i}\rho C_{i}^{\dagger}-\frac{1}{2}\{C_{i}^{\dagger}C_{i},\rho\})
\]
 and $C_{i}:=\sum_{\alpha}c(i|\alpha)\vert\alpha\rangle\langle\alpha\vert=\frac{\langle i\vert H_{I}\vert\Psi\rangle}{\langle i\vert\Psi\rangle}$.

As shown in \cite{pellegrini_diffsaut}, this equation corresponds
to the time continuous limit of repeated POVM processes (\ref{recurrho})
even if the non destruction assumption (\ref{Upointer}) is not fulfilled.

In the next section we study the long time behavior of such evolution
in the non destructive case.

\subsubsection{Long time convergence of the density matrix}

The pointer state distribution convergence indicates that,
in the long time limit, the system is in a subspace of basis $\underline{\Upsilon}$.
This information only tells us what is the limit of the elements $A_{t}(\alpha,\beta)$
when $\alpha$ or $\beta$ are not in the limit sector $\underline{\Upsilon}$.
    From the Cauchy-Schwarz theorem, $\lim_{t\to\infty}Q_{t}(\alpha)Q_{t}(\beta)=0$
implies $\lim_{t\to\infty}A_{t}(\alpha,\beta)=0$. For the elements
$A_{t}(\alpha,\beta)$ with $\alpha,\beta\in\underline{\Upsilon}$,
the limit $t\to\infty$ is yet unknown.

We decompose the operators $C_{i}$ in a sum of two hermitian operators 
\[
C_{i}=R_{i}+iS_{i}\]
with $R_{i}=\sum_{\alpha}\mathrm{Re}(c(i|\alpha))\,\vert\alpha\rangle\langle\alpha\vert$
and $S_{i}=\sum_{\alpha}\frac{1}{2}\Gamma(i|\alpha)\vert\alpha\rangle\langle\alpha\vert$.

As in the discrete time case, the density matrix evolution has to be modified by a unitary process
in order to get convergence when $t$ goes
to infinity. Let $\widetilde{U}_{t}$ be the unitary diagonal operator defined via
\begin{eqnarray*}
\widetilde{U}_{t}^{-1}\, d\widetilde{U}_{t}={-i\big(H_{s}-\sum_{i}p_{0}(i)[R_{i}(S_{i}-2\langle S_{i}\rangle_{t})-\frac{i}{2}R_i^2]\big)dt-i\sum_{i}R_{i}dX_{t}(i)}
\end{eqnarray*}
and let $\widetilde{\rho}_{t}$ be the modified density matrix
\[
\widetilde{\rho}_{t}=\widetilde{U}_{t}^{\dagger}\rho_{t}\widetilde{U}_{t}
\]
As we show below it has an almost sure limit 
\begin{eqnarray}
\lim_{t\to\infty}\widetilde{\rho}_{t}=\frac{1}{q_{0}(\underline{\Upsilon})}\mathrm{P}_{\underline{\Upsilon}}\,\rho_{0}\,\mathrm{P}_{\underline{\Upsilon}}\label{limit_rhotilde}
\end{eqnarray}
 where $\mathrm{P}_{\underline{\Upsilon}}:=\sum_{\gamma\in\underline{\Upsilon}}\vert\gamma\rangle\langle\gamma\vert$
is the projector on the subspace corresponding to the sector $\underline{\Upsilon}$.
Therefore, $\widetilde{\rho}_{\infty}$ is equivalent
to the density matrix we would have found if an initial Von Neumann measurement
had been performed on the system. The unitary operator
$\widetilde{U}_{t}$ only induces a rotation inside the limit subspace.

Recall that we only need to prove the
convergence of the $\widetilde{\rho}_{t}$ matrix elements corresponding to
two pointer states in the same sector. From the Belavkin equation
(\ref{def:belavkin_eq}) and using It\^o rules, we find the evolution equation
for $\widetilde{\rho}_{t}$:
 \[
d\widetilde{\rho}_{t}=\sum_{i}p_{0}(i)(S_{i}\widetilde{\rho}_{t}S_{i}-\frac{1}{2}\{S_{i}S_{i},\widetilde{\rho}_{t}\})dt-i\sum_{i}(\{S_{i},\widetilde{\rho}_{t}\}-2{\rm Tr}[S_{i}\widetilde{\rho}_{t}])dX_{t}(i)\]
Thus, the time evolution of matrix elements $\widetilde{A}_{t}(\alpha,\beta)$ of $\widetilde{\rho}_{t}$
with $\beta$ and $\alpha$ in the same sector is, \[
d\widetilde{A}_{t}(\alpha,\beta)=\widetilde{A}_{t}(\alpha,\beta)\sum_{i}(\Gamma(i|\alpha)-\langle\Gamma(i)\rangle_{t})dX_{t}(i)\]
Noticing that ${Q_{t}(\alpha)}{d\widetilde{A}_{t}(\alpha,\beta)}={\widetilde{A}_{t}(\alpha,\beta)} {dQ_{t}(\alpha)}$ and repeating the discussion of section \ref{sssec:conv_conttime_distrib}, we get
\[
\widetilde{A}_{\infty}(\alpha,\beta)=\left\{ \begin{array}{ll}
\frac{A_{0}(\alpha,\beta)}{q_{0}(\underline{\Upsilon})} & \mbox{if }\alpha\mbox{ and }\beta\in\underline{\Upsilon}\\
0 & \mbox{else}\end{array}\right.\]
 This proves the limit (\ref{limit_rhotilde}).

\subsubsection{Extension to different partial measurement methods}

We can extend our results to cases where different partial measurement methods are
used. Once again we limit ourselves to time independent random
protocols. The density matrix evolution is modified as follows:
\[
 d\rho_{t}=L(\rho_{t})\, dt +\sum_{(o,i)\in E}{\cal D}_{(o,i)}(\rho_t)\, dX_{t}(o,i)
 \]
with
 \[ 
 {\cal D}_{(o,i)}(\rho_t)= -i\big(C_{i}^{(o)}\rho_{t}-\rho_{t}{C_{i}^{(o)}}^{\dagger}-\rho_{t}\,{\rm Tr}[C_{i}^{(o)}\rho_{t}-\rho_{t}{C_{i}^{(o)}}^{\dagger}]\big)
 \]
 where $C_{i}^{(o)}=\frac{\langle i\vert H_{I}^{(o)}\vert\Psi_{o}\rangle}{\langle i\vert\Psi_{o}\rangle}$
and \[
L(\rho_t)=-i[H_{s},\rho_t]+\sum_{(o,i)\in E}c(o)\;p_{0}^{o}(i)\big(C_{i}^{(o)}\rho_{t}{C_{i}^{(o)}}^{\dagger}-\frac{1}{2}\{{C_{i}^{(o)}}^{\dagger}{C_{i}^{(o)}},\rho_{t}\}\big)\]
As before $c(o)$ is the probability of using measurement method $o$. 
The limit density matrix can be analyzed as above: we obtain 
identical convergence statements once the density matrix has been rotated using an appropriate unitary $\widetilde{U}_t$.


\vskip 1.0 truecm \textbf{Acknowledgements}: This work was in part
supported by ANR contract ANR-2010-BLANC-0414. T.B. thanks Clement
Pellegrini for helpful discussions on the continuous time limit. 

\vfill \eject

\appendix

\section{Details for mutual singularity}
\label{sec:details}

We prove that if $o$ is recurrent but $(o,i)$ is not then
$\sum_{\gamma} Q(\gamma)p^o(i|\gamma)=0$. 

Observe that under ${\mathbb P}_{\alpha}$ we have the Markov property
\begin{equation}
{\mathbb E}_{\alpha}({\mathbb 1}_{I_n=i}|{\mathcal F}_{n-1})=p^{O_n}(i|\alpha)
\label{eq:markov}
\end{equation}
It says that, under ${\mathbb
  P}_{\alpha}$, $O_n=o$ the next measurement outcome is $i\in \text{spec} (o)$ with
probability $p^{o}(i|\alpha)$ independently of what happened before.

Now assume that under ${\mathbb P}_{\alpha}$ measurement method $o$ is recurrent
with probability 1. Take $0\leq T_1 < T_2 < \cdots <T_k \cdots$ to be the times
when the measurement method is $o$. We show, using the strong Markov property,
that $I_{T_1},I_{T_2},\cdots$ are independent identically distributed random
variables with distribution $p^{o}(\cdot |\alpha)$. This is quite natural: the
functions $d_n$ help choosing the measurement method, but they do not influence
the measurement result.

Indeed, note first that the above statement is trivial
when there is only one measurement method, because then there is no need to
invoke stopping times and the strong Markov property. In the general case, note
the slight mismatch with usual notations: $\{T_k\leq n\}$
is in fact ${\mathcal F}_{n-1}$ measurable, so it is natural to write ${\mathcal
  F}_{T_k-1}$ for the algebra associated to the stopping time $T_k$. Then write
\begin{eqnarray*}
{\mathbb E}_{\alpha}({\mathbb 1}_{I_{T_1}=i_1}\cdots {\mathbb 1}_{I_{T_k}=i_k}|{\mathcal
  F}_{T_k-1}) & =  & {\mathbb 1}_{I_{T_1}=i_1}\cdots {\mathbb
  1}_{I_{T_{k-1}}=i_{k-1}}{\mathbb E}_{\alpha}({\mathbb
    1}_{I_{T_k}=i_k}|{\mathcal F}_{T_k-1})\\
& = & {\mathbb 1}_{I_{T_1}=i_1}\cdots {\mathbb
  1}_{I_{T_{k-1}}=i_{k-1}}p^{o}(i_k|\alpha)
\end{eqnarray*}
One can go on to condition with respect to ${\mathcal
  F}_{T_{k-1}-1}$, $\cdots$ until one finds the plain expectation 
\[
{\mathbb E}_{\alpha}({\mathbb 1}_{I_{T_1}=i_1}\cdots {\mathbb 1}_{I_{T_k}=i_k})=p^{o}(i_1|\alpha)\cdots p^{o}(i_k|\alpha).
\]

As a consequence, for any $\alpha$ such that $o$ is recurrent under ${\mathbb
  P}_{\alpha}$: 

-- either $p^{o}(i|\alpha)>0$ and with ${\mathbb P}_{\alpha}$-probability $1$
the outcome $i$ appears infinitely many times in the sequence
$I_{T_1},I_{T_2},\cdots$, i.e. $(o,i)$ is recurrent with probability $1$,

-- or $p^{o}(i|\alpha)=0$ and  $i$ never appears in the sequence
$I_{T_1},I_{T_2},\cdots$, i.e. $(o,i)$ never appears. 

Now assume that $o$ is recurrent under all ${\mathbb P}_{\alpha}$'s. The above
implies immediately that the probability under ${\mathbb P}$ that $(o,i)$ is
non-recurrent (this event is denoted by $\tilde{A}_{(o,i)}$) is given by
$\sum_{\gamma,p^{o}(i|\gamma)=0} q_{0}(\gamma)$.

If $p^{o}(i|\beta)=0$ then $Q(\beta)=Q(\beta){\mathbb 1}_{\tilde{A}_{(o,i)}}$
because by the recursion relation $Q_n(\beta)=0$ whenever $(o,i)$ has shown up
before time $n$.  So
\[ {\mathbb E}(Q(\beta)|\tilde{A}_{(o,i)})=\frac{{\mathbb
    E}(Q(\beta))}{{\mathbb E}({\mathbb
    1}_{\tilde{A}_{(o,i)}})}=\frac{q_{0}(\beta)}{\sum_{\gamma,p^{o}(i|\gamma)=0}
    q_{0}(\gamma)}
\]
which implies that 
\[ {\mathbb E}(\sum_{\gamma,p^{o}(i|\gamma)=0}Q(\gamma)|\tilde{A}_{(o,i)})=1
\] 
Hence, conditional on $\tilde{A}_{(o,i)}$, the $Q(\alpha)$'s for which
$p^{o}(i|\alpha)>0$ have to vanish. Equivalently, $Q(\gamma)p^o(i|\gamma)=0$ for
each $\gamma$ and $\sum_{\gamma} Q(\gamma)p^o(i|\gamma)=0$,
which was to be proved.

\section{Proof of existence of a continuous time limit}
\label{app:conv}

We first put the notion of continuous time limit in context.

Let $V$ be the vector space $C_{0}({\mathbb{R}}^{+},{\mathbb{R}}^{I})$
of continuous functions $f_{\cdot}$ from ${\mathbb{R}}^{+}$ to ${\mathbb{R}}^{I}$
such that $f_{0}=0\in{\mathbb{R}}^{I}$. For each $\delta>0$, and
each $\omega\in\Omega$ we define a function $W_{t}^{(\delta)}(i)$
on ${\mathbb{R}}^{+}$ by linear interpolation of $W_{t}^{(\delta)}(i):=\sqrt{\delta}(N_{t/\delta}(i)-p_{0}(i)t/\delta)$
if $t/\delta$ is an integer. Explicitly, for $t\in[\delta n,\delta(n+1)]$
\[
W_{t}^{(\delta)}(i)=\sqrt{\delta}\left((n+1-t/\delta)N_{n}(i)+(t/\delta-n)N_{n+1}(i)-p_{0}(i)t/\delta\right).\]

For every $\omega\in\Omega$ the function $W_{t}^{(\delta)}(i)$ is
continuous for $t\in{\mathbb{R}}^{+}$. So we have a map $W^{(\delta)}:\Omega\rightarrow V$.
But, as already pointed out before, there is no hope that, for a fixed
$\omega\in\Omega$, $W_{t}^{(\delta)}(i)$ has a limit when $\delta\rightarrow0^{+}$.
The only clear fact is that for a fixed $t$, the central limit theorem
ensures that the distribution of $W_{t}^{(\delta)}(i)$ under each
${\mathbb{P}}_{\alpha}$ has a Gaussian limit when $\delta\rightarrow0^{+}$.
Note that this observation fixes the scaling $\sqrt{\delta}$
as the only one possible.

But if we are interested in convergence as a process, a deeper approach
is needed. If we endow $V$ with the topology of uniform convergence
on compact sets ${\mathcal{T}}(V)$ and with the corresponding Borel
$\sigma$-algebra ${\mathcal{B}}(V)$, we can show that the map $W^{(\delta)}$
is measurable from $(\Omega,{\mathcal{F}})$ to $(V,{\mathcal{B}}(V))$.
This is not difficult, because by a classical result, ${\mathcal{B}}(V)$
is the smallest $\sigma$-algebra on $V$ containing the family of
sets \[
B^{t,i,a}:=\{f\in C_{0}({\mathbb{R}}^{+},{\mathbb{R}}^{I}),a<f_{t}(i)\}\]
 indexed by $t\in]0,+\infty[$, $i\in I$ and $a\in{\mathbb{R}}$.
It is plain that the inverse image of $B^{t,i,a}$ under $W^{(\delta)}$
is in ${\mathcal{F}}_{n}$ whenever $n>t/\delta$. As the filtration
on $\Omega$ is exactly the one making the $N_{n}$ ${\mathcal{F}}_{n}$-measurable,
the appropriate filtration on $C({\mathbb{R}}^{+},{\mathbb{R}}^{I})$
should be the natural one, the smallest making the canonical process
adapted%
\footnote{Of course, as long as $\delta>0$, $W_{t}$ looks a bit forward in
the future, as it involves $N_{n+1}$ for $t\in[\delta n,\delta(n+1)]$,
but for continuous process this does not matter.%
}. We denote it by ${\mathcal{G}}_{t}$.

Then any probability measure on $(\Omega,{\mathcal{F}})$ induces
via $W^{(\delta)}$ a probability measure on $(V,{\mathcal{B}}(V))$.
Note that, via (\ref{eq:pdelta}), the measures we defined previously
on $(\Omega)$ depend on $\delta$, and to be explicit we write ${\mathbb{P}}^{(\delta)}$,
${\mathbb{E}}^{(\delta)}$, etc. to stress this fact. Let $\mu_{\alpha}(\delta)$
be the image measure of ${\mathbb{P}}_{\alpha}^{(\delta)}$ pushed
forward by $W^{(\delta)}$ on $(V,{\mathcal{B}}(V))$. As $(V,{\mathcal{T}}(V))$
is a so-called Polish space, there is a nice notion of convergence
for measures on it, called {}``weak convergence'' of measures, and
we could ask if the $\mu_{\alpha}(\delta)$'s converge weakly to some
probability measure on $(V,{\mathcal{B}}(V))$ (and then, so would
$\mu(\delta)=\sum_{\alpha}q_{0}(\alpha)\mu_{\alpha}(\delta))$. Note
that despite its name, weak convergence is strong enough to ensure
the convergence of the expectations of rather general functionals,
so we could hope to control the $Q$'s continuous time limit as well, because they are nice functionals of the counting process.

It is usually in this context that continuous time limits have a meaning.
In this setting, there are a number of theorems, called functional
central limit theorems, or Donsker invariance principles, that express
the continuous time limit of random walks (with independent increments)
in terms of Brownian motions in fine details. Alas, though under each
${\mathbb{P}}_{\alpha}$, $N_{n}(i)$ is a random walk with independent
increments, the theorems we are aware do not apply immediately. The
problem is that the $\delta$ dependence is not only in $W$, but
also in the ${\mathbb{P}}_{\alpha}$'s. This problems would show up
even more dramatically to deal with the $Q$'s convergence 
and there relationships with $W$. While we think these are purely
technical details in our case, we shall not try to deal with them:
we shall instead rely on a weaker and slightly less natural notion
of continuous time limit that will suffice for our purposes. To say things in
more mathematical terms: we shall content ourselves with proving that the
joint finite dimensional distributions of $W$'s and $Q$'s converge to joint
finite dimensional distributions of continuous processes we can identify
explicitly, but we do not embark on the more technical task of proving tightness. 

\vspace{0.3cm}

We now turn to our explicit approach of the continuous time limit.

 From the characteristic function (\ref{eq:charfunc}) we obtain easily
that \begin{eqnarray}
\lim_{\delta\rightarrow0^{+}}{\mathbb{E}}^{(\delta)}\left(e^{\sum_{l=1}^{k}\sum_{i}\lambda_{l}(i)(W_{t_{l}}^{(\delta)}(i)-W_{t_{l-1}}^{(\delta)}(i))}\right) & =\nonumber \\
 &  & \hspace{-4cm}\left(\sum_{\alpha}q_{0}(\alpha)e^{\sum_{l=1}^{k}(t_{l}-t_{l-1})\sum_{i}\lambda_{l}(i)p_{0}(i)\Gamma(i|\alpha)}\right)\;\times\nonumber \\
 &  & \hspace{-4cm}e^{\frac{1}{2}\sum_{l=1}^{k}(t_{l}-t_{l-1})(\sum_{i}p_{0}(i)\lambda_{l}(i)^{2}-(\sum_{i}p_{0}(i)\lambda_{l}(i))^{2})}\label{eq:charfunc2}\end{eqnarray}
 for $k\geq1$, arbitrary non-decreasing sequences $0=t_{0}\leq t_{1}\leq\cdots\leq t_{k}$
of length $k$ and arbitrary (complex) $\lambda_{l}(i)$'s.

By a standard theorem on characteristic functions, we have thus proved
that the $W_{t}^{(\delta)}(i)$ finite marginals (under each
${\mathbb{P}}_{\alpha}^{(\delta)}$ and under ${\mathbb{P}}^{(\delta)}$)
have a limit for $\delta\rightarrow0^{+}$. This is much weaker than
what weak convergence of measures would ensure. It is enough to ensure
that the limit marginals satisfy the Kolmogorov consistency criterion,
but it does not guarantee that it is possible to concentrate the corresponding
process on $C_{0}({\mathbb{R}}^{+},{\mathbb{R}}^{I})$. However in
the case at hand, we can bypass this problem because of the simple
form of the result, which is Gaussian for each $\alpha$.

Let $\nu$ be the Wiener measure of a standard Brownian motion on
$C_{0}({\mathbb{R}}^{+},{\mathbb{R}}^{I})$. The linear map from ${\mathbb{R}}^{I}$
to itself defined by $$y(i):=\sqrt{p_{0}(i)}(x(i)-\sqrt{p_{0}(i)}\sum_{j}\sqrt{p_{0}(j)}x(j))$$
induces a map from $C_{0}({\mathbb{R}}^{+},{\mathbb{R}}^{I})$ to
itself. Let $\mu^{0}$ be the image measure of $\nu$ under this map.
It is easily seen that under this law, the canonical process $W$
on $C_{0}({\mathbb{R}}^{+},{\mathbb{R}}^{I})$ satisfies

\begin{eqnarray}
{\mathbb{E}}^{\mu^{0}}\left(e^{\sum_{l=1}^{k}\sum_{i}\lambda_{l}(i)(W_{t_{l}}(i)-W_{t_{l-1}}(i))}\right) & =\nonumber \\
 &  & \hspace{-4cm}e^{\frac{1}{2}\sum_{l=1}^{k}(t_{l}-t_{l-1})(\sum_{i}p_{0}(i)\lambda_{l}(i)^{2}-(\sum_{i}p_{0}(i)\lambda_{l}(i))^{2})}\label{eq:charfunc3}\end{eqnarray}
 for $k\geq1$, arbitrary non-decreasing sequences $0=t_{0}\leq t_{1}\leq\cdots\leq t_{k}$
of length $k$ and arbitrary (complex) $\lambda_{l}(i)$'s.

As a consequence, It\^o's formula holds with \[
dW_{t}(i)dW_{t}(j)=dt\,\big(\delta_{i,j}\, p_{0}(i)-p_{0}(i)p_{0}(j)\big).\]

Our aim is to use Girsanov's theorem to deform the measure and go
from the right-hand side in (\ref{eq:charfunc3}) to the right-hand
in (\ref{eq:charfunc2}). If the process $U_{t}$ with values in ${\mathbb{R}}^{I}$
is adapted and satisfies some further technical integrability conditions,
\[
M_{t}^{U}:=e^{\int_{0}^{t}\sum_{i}U_{s}(i)dW_{s}(i)-\frac{1}{2}\int_{0}^{t}(\sum_{i}p_{0}(i)U_{s}(i)^{2}-(\sum_{i}p_{0}(i)U_{s}(i))^{2})ds}\]
 is a martingale. Moreover, by It\^o's formula, $dM^U_{t}=M_{t}^{U}\sum_{i}U_{s}(i)dW_{s}(i)$.
Then, by Girsanov's theorem, for any $T>0$, under the measure $d\mu_{T}^{U}:=M_{T}^{U}d\mu^{0}$
on $C_{0}([0,T],{\mathbb{R}}^{I})$, the process $W_{t}(i)-p_{0}(i)\int_{0}^{t}(U_{s}(i)-\sum_{j}p_{0}(j)U_{s}(j))ds$,
$t\in[0,T]$, has the same law as $W_{t}(i)$, $t\in[0,T]$, under
$d\mu^{0}$.

Note that for $t\leq T$, $M_{t}^{U}$ is a Radon-Nikodym derivative.
\[
M_{t}^{U}:=\left[\frac{d\mu_{T}^{U}}{d\mu^{0}}\right]_{t}={\mathbb{E}}^{\mu^{0}}\left(\frac{d\mu_{T}^{U}}{d\mu^{0}}\Big|{\mathcal{G}}_{t}\right),\]

In general this construction cannot work for infinite $T$, because
$\mu_{T}^{U}$ and $\mu^{0}$ become singular. However, $T$ plays
only a dummy role : for $T'\leq T$, $d\mu_{T}^{U}$ and $d\mu_{T'}^{U}$
coincide on ${\mathcal{G}}_{T'}$. So it is only a slight abuse, which
lightens notations a bit, to write $d\mu^{U}$ for $d\mu_{T}^{U}$
and \[
M_{t}^{U}=\left[\frac{d\mu^{U}}{d\mu^{0}}\right]_{t}={\mathbb{E}}^{\mu^{0}}\left(\frac{d\mu^{U}}{d\mu^{0}}\Big|{\mathcal{G}}_{t}\right).\]

For the special choice $U_{t}(i):=\Gamma(i|\alpha)$, using $\sum_{i}p_{0}(i)\Gamma(i|\alpha)=0$,
we compute \[
M_{t}(\alpha):=e^{\sum_{i}\Gamma(i|\alpha)W_{t}(i)-\frac{t}{2}\sum_{i}p_{0}(i)\Gamma(i|\alpha)^{2}}\]
 which is certainly a martingale, and we obtain for every $T$ a measure
$d\mu_{\alpha}$ on $C_{0}([0,T],{\mathbb{R}}^{I})$ such that \begin{eqnarray}
{\mathbb{E}}^{\mu_{\alpha}}\left(e^{\sum_{l=1}^{k}\sum_{i}\lambda_{l}(i)(W_{t_{l}}(i)-W_{t_{l-1}}(i))}\right) & =\nonumber \\
 &  & \hspace{-8cm}e^{\sum_{l=1}^{k}(t_{l}-t_{l-1})\sum_{i}\lambda_{l}(i)p_{0}(i)\Gamma(i|\alpha)}e^{\frac{1}{2}\sum_{l=1}^{k}(t_{l}-t_{l-1})(\sum_{i}p_{0}(i)\lambda_{l}(i)^{2}-(\sum_{i}p_{0}(i)\lambda_{l}(i))^{2})}\end{eqnarray}
 for $k\geq1$, arbitrary non-decreasing sequences $0=t_{0}\leq t_{1}\leq\cdots\leq t_{k}\leq T$
of length $k$ and arbitrary (complex) $\lambda_{l}(i)$'s.

Finally, setting $M_{t}:=\sum_{\alpha}q_{0}(\alpha)M_{t}(\alpha)$
(trivially a martingale again) and $d\mu:=M_{t}d\mu^{0}$ we obtain
for every $T$ a measure $d\mu$ on $C_{0}([0,T],{\mathbb{R}}^{I})$
such that \begin{eqnarray}
{\mathbb{E}}^{\mu}\left(e^{\sum_{l=1}^{k}\sum_{i}\lambda_{l}(i)(W_{t_{l}}(i)-W_{t_{l-1}}(i))}\right) & =\nonumber \\
 &  & \hspace{-4cm}\left(\sum_{\alpha}q_{0}(\alpha)e^{\sum_{l=1}^{k}(t_{l}-t_{l-1})\sum_{i}\lambda_{l}(i)p_{0}(i)\Gamma(i|\alpha)}\right)\;\times\nonumber \\
 &  & \hspace{-4cm}e^{\frac{1}{2}\sum_{l=1}^{k}(t_{l}-t_{l-1})(\sum_{i}p_{0}(i)\lambda_{l}(i)^{2}-(\sum_{i}p_{0}(i)\lambda_{l}(i))^{2})}\label{eq:charfunc5}\end{eqnarray}
 for $k\geq1$, arbitrary non-decreasing sequences $0=t_{0}\leq t_{1}\leq\cdots\leq t_{k}\leq T$
of length $k$ and arbitrary (complex) $\lambda_{l}(i)$'s. So we
have found the continuous time limit of the counting process as the
canonical process on $C_{0}([0,T],{\mathbb{R}}^{I})$ with the measure
$\mu$.

It remains to deal with the continuous time limit of the $Q_{n}$'s.

We note that by the chain rule $\frac{M_{t}(\alpha)}{M_{t}}=\left[\frac{d\mu(\alpha)}{d\mu}\right]_{t}={\mathbb{E}}^{\mu}\left(\frac{d\mu(\alpha)}{d\mu}\Big|{\mathcal{G}}_{t}\right)$
is the Radon-Nikodym derivative of $\mu(\alpha)$ with respect to
$\mu$. So with some memory of what happened in the discrete case,
it is natural to define $Q_{t}(\alpha):=q_{0}(\alpha)\frac{M_{t}(\alpha)}{M_{t}}$.

Using the explicit formula for the $Q_{n}$'s in terms of the counting
processes, we define $Q_{t}^{(\delta)}$ by an interpolation procedure:
\[
Q_{t}^{(\delta)}(\alpha):=q_{0}(\alpha)\frac{\prod_{i}p(i|\alpha)^{W_{t}^{(\delta)}(i)/\sqrt{\delta}+p_{0}(i)t/\delta}}{\sum_{\beta}q_{0}(\beta)\prod_{i}p(i|\beta)^{W_{t}^{(\delta)}(i)/\sqrt{\delta}+p_{0}(i)t/\delta}},\]
 so that if $t/\delta=n$, an integer, $Q_{t}^{(\delta)}(\alpha)=Q_{n}(\alpha)$.
Note that in this formula the $p(i|\beta)$'s depend implicitly on
$\delta$ via (\ref{eq:pdelta}).

One can prove that the joint finite dimensional distributions of the
processes $(W_{t}^{(\delta)},Q_{t}^{(\delta)})$ under ${\mathbb{P}}^{(\delta)}$
have limits when $\delta\rightarrow0^{+}$ and that these limit are
nothing but the joint finite dimensional distributions of the processes
$(W_{t},Q_{t})$ under $\mu$. In this precise sense, the $Q_{n}$'s continuous
time limit is deciphered.

This result is really no big surprise, but to prove it we have to
rely on an \textit{ad hoc} trick and an explicit elementary but tedious
computation. The details are neither illuminating nor elegant so we
omit them.

When $\delta$ is small enough, all non-empty sets in ${\mathcal{F}}_{n}$
have strictly positive measure, so that if $q_{0}(\alpha)>0$ the
same is true for $Q_{n}(\alpha)$ for all $n$'s. As furthermore $\sum_{\alpha}Q_{n}(\alpha)=1$
for all $n$, all the information on the $Q_{n}$'s (joint) laws is embodied in the joint laws of ratios of $Q_{n}$'s.

As these ratios have a simple product structure in terms of the counting
processes, the explicit computation of \[
{\mathbb{E}}\left(e^{\sum_{l=1}^{k}\sum_{i}\lambda_{l}(i)(N_{n_{l}}(i)-N_{n_{l-1}}(i))}\prod_{l=1}^{k}\prod_{\alpha,\beta}\left(\frac{Q_{n_{l}}(\alpha)}{Q_{n_{l}}(\beta)}\right)^{\eta_{l}(\alpha,\beta)}\right)\]
 for $k\geq1$, arbitrary non-decreasing sequences of integers $0=n_{0}\leq n_{1}\leq\cdots\leq n_{k}$
of length $k$, and arbitrary (complex) $\lambda_{l}(i)$'s and $\eta_{l}(\alpha,\beta)$'s
is in some sense a special case of (\ref{eq:charfunc}).

The same remark applies to computations involving ratios of $Q_{t}$'s.

This allows to compute explicitly that \begin{eqnarray}
\lim_{\delta\rightarrow0^{+}}{\mathbb{E}}^{(\delta)}\left(e^{\sum_{l=1}^{k}\sum_{i}\lambda_{l}(i)(W_{t_{l}}^{(\delta)}(i)-W_{t_{l-1}}^{(\delta)}(i))}\prod_{l=1}^{k}\prod_{\alpha,\beta}\left(\frac{Q_{t_{l}}^{(\delta)}(\alpha)}{Q_{t_{l}}^{(\delta)}(\beta)}\right)^{\eta_{l}(\alpha,\beta)}\right) & =\nonumber \\
 &  & \hspace{-10.5cm}{\mathbb{E}}^{\mu}\left(e^{\sum_{l=1}^{k}\sum_{i}\lambda_{l}(i)(W_{t_{l}}(i)-W_{t_{l-1}}(i))}\prod_{l=1}^{k}\prod_{\alpha,\beta}\left(\frac{Q_{t_{l}}(\alpha)}{Q_{t_{l}}(\beta)}\right)^{\eta_{l}(\alpha,\beta)}\right)\end{eqnarray}
 for $k\geq1$, arbitrary non-decreasing sequences $0=t_{0}\leq t_{1}\leq\cdots\leq t_{k}$
of length $k$, and arbitrary (complex) $\lambda_{l}(i)$'s and $\eta_{l}(\alpha,\beta)$'s.

As such a mixture of Laplace and Mellin transforms characterizes the
distributions completely, this concludes the existence of a natural
continuous time limit.

\section{Details on the continuous time limit with different partial measurement methods}
\label{app:detailB}

We use the linear interpolation of appendix \ref{app:conv}
on $W_{t}^{(\delta)}(o,i):=\sqrt{\delta}(N_{t/\delta}(o,i)-c(o)p_{0}^{o}(i)t/\delta)$
if $t/\delta$ is an integer. Explicitly for $t\in[\delta n,\delta(n+1)]$,
\[
W_{t}^{(\delta)}(o,i)=\sqrt{\delta}\left((n+1-t/\delta)N_{n}(o,i)
+(t/\delta-n)N_{n+1}(o,i)-c(o)p_{0}^{o}(i)t/\delta\right).
\]

We remind that $E=\bigcup_{o\in\mathcal{O}}\{o\}\otimes\spec(o)$
is the set of all possible measurement methods and outcomes. We expect the limit time continuous process to live on the vector space of continuous function from $\mathbb{R}^{+}$ to $\mathbb{R}^{E}$.

Compare to previous sections, the main changes are
in the correlation functions calculations. Thanks to the measurement
method distribution time and realization independency, we find:
\begin{eqnarray}
{\mathbb{E}}\left(e^{\sum_{l=1}^{k}\sum_{(o,i)\in E}\lambda_{l}(o,i)(N_{n_{l}}(o,i)-N_{n_{l-1}}(o,i))}\right) & =\nonumber \\
 &  & \hspace{-5cm}\sum_{\alpha}q_{0}(\alpha)\prod_{l=1}^{k}\left(\sum_{(o,i)\in E}e^{\lambda_{l}(o,i)}c(o)p^{o}(i|\alpha)\right)^{n_{l}-n_{l-1}}
 \end{eqnarray}
 for $k\geq1$, arbitrary non-decreasing sequences of integers $0=n_{0}\leq n_{1}\leq\cdots\leq n_{k}$
of length $k$, and arbitrary (complex) $\lambda_{l}(o,i)$'s.

As a consequence, in the limit $\delta\to0^{+}$, \begin{eqnarray}
\lim_{\delta\rightarrow0^{+}}{\mathbb{E}}^{(\delta)}\left(e^{\sum_{l=1}^{k}\sum_{(o,i)\in E}\lambda_{l}(o,i)(W_{t_{l}}^{(\delta)}(o,i)-W_{t_{l-1}}^{(\delta)}(o,i))}\right) & =\nonumber \\
 &  & \hspace{-9cm}\left(\sum_{\alpha}q_{0}(\alpha)e^{\sum_{l=1}^{k}(t_{l}-t_{l-1})\sum_{(o,i)\in E}\lambda_{l}(o,i)c(o)p_{0}^{o}(i)\Gamma^{(o)}(i|\alpha)}\right)\;\times\nonumber \\
 &  & \hspace{-9cm}e^{\frac{1}{2}\sum_{l=1}^{k}(t_{l}-t_{l-1})(\sum_{(o,i)\in E}c(o)p_{0}^{o}(i)\lambda_{l}(o,i)^{2}-(\sum_{(o,i)\in E}c(o)p_{0}^{o}(i)\lambda_{l}(o,i))^{2})}\end{eqnarray}
 for $k\geq1$, arbitrary non-decreasing sequences $0=t_{0}\leq t_{1}\leq\cdots\leq t_{k}$
of length $k$ and arbitrary (complex) $\lambda_{l}(o,i)$'s and with
$\Gamma^{(o)}(i|\alpha)=2\mathrm{Im}\left(\frac{\langle i\vert H_{\alpha}^{(o)}\vert\Psi^{(o)}\rangle}{\langle i\vert\Psi^{(o)}\rangle}\right)$.
Then each $W_{t}^{(\delta)}(o,i)$ under $\mu^{(\delta)}$ converges toward a process $W_{t}(o,i)$ under $\mu$.

The demonstration is then as in previous section except for notational differences which keep track of the $o$-dependency of $W_{t}(o,i)$. As in the previous section, the measure $\mu^0$ is defined as the push-forward measure of $\nu$ on $\mathbb{R}^E$ under the linear map
\[
y(o,i):=\sqrt{{c(o)p_{0}^{o}(i)}}(\, x(o,i)-\sqrt{c(o)p_{0}^{o}(i)}\sum_{(o',j)\in E}\sqrt{c(o')p_{0}^{o'}(j)}\, x(o',j)
\]
The martingale $M_{t}$ is defined by 
 \[
M_{t}=\sum_{\alpha}q_{0}(\alpha)e^{\sum_{(o,i)\in E}\Gamma^{(o)}(i|\alpha)W_{t}(o,i)-\frac{t}{2}\sum_{(o,i)\in E}c(o)p_{0}^{o}(i)\Gamma^{(o)}(i|\alpha)^{2}}
\]
The measure $\mu$ is defined via Girsanov's transformation: $\mathbb{E}^{\mu}(\cdot)=\mathbb{E}^{\mu^{0}}(M_{t}\cdot)$.

\section{Derivation of the density matrix evolution}
\label{app:density}
Let us derive the density matrix continuous time limit.
Recall that at time $n$ its elements are functions of the counting
processes 
\begin{eqnarray*}
A_{n}(\alpha,\beta)=\frac{A_{0}(\alpha,\beta)\prod_{i}(M(i|\alpha)M(i|\beta)^{\star})^{N_{n}(i)}}{\sum_{\gamma}q_{0}(\gamma)\prod_{i}p(i|\gamma)^{N_{n}(i)}}
\end{eqnarray*}
 from this expression we define time continuous processes 
 \begin{eqnarray*}
A_{t}^{(\delta)}(\alpha,\beta)=\frac{A_{0}(\alpha,\beta)\prod_{i}(M(i|\alpha)M(i|\beta)^{\star})^{W_{t}^{(\delta)}(i)/\sqrt{\delta}+p_{0}(i)\frac{t}{\delta}}}{\sum_{\gamma}q_{0}(\gamma)\prod_{i}p(i|\gamma)^{W_{t}^{(\delta)}(i)/\sqrt{\delta}+p_{0}(i)\frac{t}{\delta}}}\end{eqnarray*}
 equal to $A_{n}(\alpha,\beta)$ if $t/\delta=n$ is an integer. The $M(i|\alpha)$'s depend explicitly on $\delta$ via
 \[
M(i|\alpha)=\langle i\vert e^{-i\delta(E_{\alpha}\mathbb{I}+H_{p}+\frac{1}{\sqrt{\delta}}H_{\alpha})}\vert\Psi\rangle
\]
 We rewrite the products over the partial measurement results as exponentials
of sums \[
A_{t}^{(\delta)}(\alpha,\beta)=\frac{A_{0}(\alpha,\beta)e^{\sum_{i}\ln(M(i|\alpha)M(i|\beta)^{\star}/p_{0}(i))(W_{t}^{(\delta)}(i)/\sqrt{\delta}+p_{0}(i)\frac{t}{\delta})}}{\sum_{\gamma}q_{0}(\gamma)e^{\sum_{i}\ln(p(i|\gamma)/p_{0}(i))(W_{t}^{(\delta)}(i)/\sqrt{\delta}+p_{0}(i)\frac{t}{\delta})}}\]
A detailed analysis of the limit $\delta\to0^+$ would require to perform the same study as in section \ref{app:conv}. However, at this stage we are confident enough to state that we can safely shortcut a few steps and use directly that $W_{t}^{(\delta)}(i)$ converge. Using $\langle \Psi| H_I|\Psi\rangle=0$, and the identity $\sum_{i}p_0(i)\frac{\langle i\vert H_\alpha^2 \vert \Psi \rangle}{\langle \Psi \vert i \rangle}=\sum_{i}p_{0}(i)|c(i|\alpha)|^{2}$, we obtain
\begin{eqnarray*}
\lim_{\delta\to0}\sum_{i}\ln(M(i|\alpha)M(i|\beta)^{\star}/p_{0}(i))(W_{t}^{(\delta)}(i)/\sqrt{\delta}+p_{0}(i)\frac{t}{\delta})\\
 & \hspace{-7cm}= l(\alpha,\beta)t-i\sum_{i}(c(i|\alpha)-c(i|\beta)^{\star})W_{t}(i)\end{eqnarray*}
where the limit as to be understood as the limit of any finite dimensional correlation functions. Therefore 
\begin{align*}
\lim_{\delta\to0}A_{t}^{(\delta)}(\alpha,\beta)= A_{t}(\alpha,\beta)\end{align*}
 with $A_{t}(\alpha,\beta)$ defined in (\ref{def:A_t}).
It is then a simple matter, using It\^o rules for $W_{t}(i)$, to derive
the Belavkin equation (\ref{def:belavkin_eq}) for the density
matrix $\rho_{t}=\sum_{\alpha,\beta}A_{t}(\alpha,\beta)\vert\alpha\rangle\langle\beta\vert$.

\end{document}